\documentclass[aip,amsmath,amssymb,preprint]{revtex4-1}
\pdfoutput=1 
\usepackage{graphicx}
\usepackage{dcolumn}
\usepackage{bm}
\usepackage[utf8]{inputenc}
\usepackage[T1]{fontenc}
\usepackage{mathptmx} 
\usepackage{etoolbox}
\usepackage{amsfonts}
\usepackage{graphicx}  
\usepackage{epstopdf,epsfig}
\usepackage{textgreek}
\usepackage{amsmath}
\usepackage{mathtools}
\usepackage{subfigure}
\usepackage{float}
\usepackage{epsfig}
\usepackage{amssymb}
\usepackage{psfrag}
\usepackage{natbib}
\usepackage{stmaryrd}
\usepackage{picture}
\usepackage{float}
\usepackage[colorlinks=true,breaklinks=true,linkcolor=blue]{hyperref}
\usepackage{mathrsfs}
\usepackage[scr=rsfs,cal=boondox]{mathalfa}

\newcommand{\be}{\begin{equation}}
\newcommand{\ee}{\end{equation}}
\newcommand{\bef}{\begin{figure}}
\newcommand{\ef}{\end{figure}}
\newcommand{\bear}{\begin{eqnarray}}
\newcommand{\ear}{\end{eqnarray}}
\newcommand{\barr}{\begin{array}}
\newcommand{\earr}{\end{array}}

\newcommand{\nn}{\nonumber}
\newcommand{\td}[1]{\tilde{#1}}
\newcommand{\f}{\frac}

\newcommand{\del}{\partial}
\newcommand{\bs}{\boldsymbol}

\newcommand{\msb}{\bm}
\usepackage{textgreek}
\usepackage{xcolor}
\usepackage{upgreek}
\usepackage{graphicx,epstopdf}

\makeatletter
\def\@email#1#2{%
 \endgroup
 \patchcmd{\titleblock@produce}
  {\frontmatter@RRAPformat}
  {\frontmatter@RRAPformat{\produce@RRAP{*#1\href{mailto:#2}{#2}}}\frontmatter@RRAPformat}
  {}{}
}%
\makeatother
\begin{document}


\title{Nonspherical oscillations of an encapsulated microbubble with interface energy under the acoustic field}
\affiliation{Department of Applied Mechanics, Indian Institute of Technology Madras, Chennai, 600036, India.}
\author{Nehal Dash}
\author{Ganesh Tamadapu}
\email{nehaldash.dash@gmail.com}

\begin{abstract}
The practical applications of gas-filled encapsulated microbubbles involve inherent nonspherical oscillations under acoustic fields. The gas-encapsulation and encapsulation-liquid interfaces significantly affect the mechanics of the bubbles, especially of smaller radii, and their consideration is vital for mimicking the experimental setting.
In this paper, we apply the interface energy model [N. Dash and G. Tamadapu, J. Fluid Mech. \textbf{932}, A$26$ $(2022)$] to examine the nonspherical oscillations of an encapsulated microbubble with a radius of $2\,\mu$m and $5\,\mu$m under an acoustic field. Using the Lagrangian energy formulation, the coupled dynamical governing equations for spherical and nonspherical modes are derived, incorporating the effects of interface energy at the interfaces, shell elasticity, and viscosity. Through a perturbation analysis based on the Krylov-Bogoliubov method of averaging, a set of first-order differential (slow-time) equations is obtained to conduct steady-state and conditional-stability analysis. The stability analysis helped in determining the excitation pressure and frequency of the acoustic field required for smaller radii bubbles to exhibit finite amplitude shape oscillations. Direct numerical simulations of the governing equations revealed that the parametrically forced even mode ($n=2$) excites even modes, while the odd modes ($n=3$) excite both even and odd modes. For smaller radii bubbles, we observe shape mode oscillations of finite non-zero amplitudes only in the presence of interface parameters. The initial size-dependent interface parameter and shell viscoelastic parameters are identified as the key parameters that play a critical role in exhibiting finite shape mode oscillations of the bubble.
\end{abstract}

\maketitle


\section{Introduction}
Encapsulated microbubbles (EBs) have emerged as contrast agents in clinical ultrasound imaging. \citep{hoff2001acoustic,Stride2003,postema2004ultrasound,Lindner2004,Klibanov2006} In addition to their established applications, research has unveiled their potential in various other fields such as microvascular imaging and blood flow, \citep{errico2015ultrafast} targeted drug delivery, \citep{1347-4065-38-5S-3014,Tsutsui2004,HERNOT20081153,HYNYNEN20081209,KOOIMAN201428} sonochemistry, \citep{Suslick1990,Blake1999} for biofilm removal, \citep{Agarwal2012,Seo2018} cavitation cleaning, \citep{Ohl2006,REUTER2017542} among others. 

In practical applications, maintaining the stability of EBs throughout the process is vital to ensure their efficacy. Different techniques have been developed to fabricate microbubbles, and enhance their controlled stability. \citep{lee2015stabilization} However, when exposed to a sufficiently intense acoustic field, the oscillating surface of the EB suspended in a fluid becomes unstable. These instabilities result in the initiation of nonspherical (shape) oscillations in the EB when the driving acoustic pressure surpasses a critical threshold. 
While the behavior of EBs inherently involves nonspherical oscillations in their applications, for the purpose of simplification in analysis, most of the studies on EBs focused on developing bubble shell models undergoing radial symmetric oscillations. \citep{DEJONG199295,DEJONG1993175,Church1995} Subsequently researchers reported the presence of nonspherical or surface mode oscillations in shell-free bubble models \citep{brenner1995bubble,doi:10.1063/1.869996} demonstrating their relevant applications in the field of sonoluminescence. However, there has always been a question of whether such observations would also be observed in shelled bubble models, given the effects of the bubble shell on the damped oscillations. \citet{1417703} and \citet{4151897} described the existence of surface modes in shelled bubbles used as contrast agents. They also conveyed that the surface modes evolve as parametric instabilities. Later, researchers investigated the parametric shape instabilities in the coated microbubbles. \citep{dollet2008nonspherical,vos2011nonspherical} Numerical models have also been developed to study the shape oscillations of the encapsulated microbubbles. \citep{liu2011numerical,tsiglifis2011parametric} \citet{liu2011numerical} used a boundary-fitted finite-volume method to model the movement of boundary of the bubble, where the shell was considered as a neo-Hookean membrane with an energy dissipation equation to capture the shell deformation. \citet{tsiglifis2011parametric} developed the numerical model to study the parametric stability and dynamic buckling of the encapsulated microbubble. \citet{tamadapu2016modeling} investigated the resonance characteristics of spherical and nonspherical modes in a thick encapsulated bubble filled with air and suspended in water, where the shell material was considered to linear viscoelastic and quasi-incompressible.

A noticeable aspect of understanding these nonspherical oscillations is to have a better understanding of the physics or behavior of a single spherical EB suspended in a viscous medium. Another important aspect is establishing an explanation for the instabilities observed in EBs, \citep{dollet2008nonspherical,vos2011nonspherical} which may lead to collapse/rupture/buckling of these EBs. \citep{doi:10.1063/1.869131,RevModPhys.74.425} With these motivations, \citet{doi:10.1121/1.4707479} proposed a reduced analytical model for bubble surface mode oscillations that takes into account the effect of shell properties on the oscillations. 
\citet{doinikov2004translational} conducted a study on the nonlinear coupling between spherical modes, translational motion, and shape modes of an oscillating bubble, with a particular focus on translational instability. They derived a set of coupled equations that described spherical oscillations of a bubble, its translational motion and shape oscillations evolving on the bubble surface. Taking this a step further, \citet{Shaw2006} studied the nonlinear interactions between the axisymmetric shape distortions, the axial translational motion, and the volume oscillations of a gas bubble in an inviscid, incompressible liquid. They represented the surface deformation by a complete set of Legendre polynomials and assumed that the deformation and translational motion are small. By using a Lagrangian energy formulation, they derived a system of equations that remained valid up to third order approximation in these interaction terms. \citet{shaw2009stability} also studied the stability and nonspherical oscillations of a bubble assuming that the bubble translation and deformation is small. Additionally, \citet{shaw2009stability} discussed the stability and nonspherical oscillations of a bubble, employing the assumption that both bubble translation and deformation are small. They used a combination of Rayleigh dissipation function and perturbation analysis to account for the effects of viscosity. In the subsequent study \citet{shaw2017nonspherical} considered the role on nonlinear shape mode interactions in bubble dynamics. They identified the parametrically identified excited shape mode together with nonlinear excitement of other shape modes, modification of the spherical mode and induced translation of the bubble. Introducing the damping effect in compressibility and viscosity, they showcased that a nonlinear coupling between parametrically forced shape mode and other modes is essential for the bubble to achieve stable oscillatory shape deformation.
Recently \citet{guedra2018bubble} followed the mathematical formulation of \citet{Shaw2006} and derived analytical solutions for weakly nonlinear shape oscillations of finite amplitude. Using perturbation methods they analysed the shape oscillations of bubbles at small amplitudes via spherical harmonics. Under the steady-state conditions, the equations yield analytical expressions of the modal amplitudes, conditionally stable and absolutely stable thresholds for shape oscillations are derived and analysed.

Lately, \citet{dash_tamadapu_2022} developed a mathematical model based on interface energy within the framework of surface continuum mechanics, \citep{steigmann,GAO201459} proposing its relevance for radial oscillations of an EB. This model highlighted the significance of interface energy at the gas-encapsulation and the encapsulation-fluid interfaces in understanding and analyzing radial oscillations of EBs. The interface energy model explained the influence of each of the interface parameters, which exerted their effects through area strain, curvature, initial size dependence, and the coupled effects of area strain and curvature. This model, for the first time, also resolved the spurious dependency of shell viscoelastic parameters on the initial size of the bubble, which remained hitherto unexplained. 
Very recently this was experimentally clarified by \citet{D3SM00871A} where they characterized the shell dilatational viscosity analysing the bubble's time domain response using ultra-high-speed microscopic imaging and optical trapping, instead of conventional bubble spectroscopy approach.
Expanding on the interface energy model, \citet{dash2022describing} provided a description of the radial dynamics of an EB with a nonlinear viscoelastic shell. The interface energy model has presented numerous possibilities for studying the behavior of EBs, especially for smaller radii bubbles. While several models have been developed to investigate the nonlinear shape oscillations of an EB in viscous medium, the investigation of nonspherical oscillations of a smaller radii EB under acoustic field, considering the influence of interface energy at the two interfaces remains unexplored.

With this motivation, we study the spherical (volumetric) and nonspherical (shape) oscillations of a gas-filled microbubble using the interface energy model. The bubble is encapsulated with a thin shell membrane and suspended in an infinite incompressible medium. The influence of interface energy at both the gas-encapsulation and encapsulation-liquid interfaces is taken into consideration. Furthermore, the effects of elasticity and viscosity of the thin shell membrane are incorporated into the analysis. The resulting coupled governing equations for spherical and nonspherical modes of bubble oscillations, which captures the effects of interface energy and shell elasticity, are derived. The behavior of the spherical bubble in both the spherical and nonspherical modes is analyzed using direct numerical simulations, considering a reasonable set of physical and interface parameters. In order to explore the conditional stability of the EB, a perturbation analysis using Krylov-Bogoliubov averaging method is carried out. While the model developed by \citet{guedra2018bubble} focused on analyzing nonspherical oscillations in uncoated bubbles of larger radius, the present study is directed towards encapsulated bubbles with smaller radii, where the influence of interfaces becomes particularly significant. The novelty of the present study unfolds in three key aspects: (i) it introduces an interface energy-based mathematical model aimed at analyzing the nonspherical oscillations of an EB with smaller radii, (ii) the model considers the shell material as a thin viscoelastic membrane, introducing elasticity and viscosity parameters into the mathematical model, (iii) through numerical simulations and stability analysis the model highlights the importance of interface parameters in existence of shape mode oscillations in smaller radii EBs with interface effects. 

This paper is organized as follows. Section~\ref{sec:kindef} introduces the kinematics of deformation of the EB. Section~\ref{sec:mathmodel} presents the mathematical model formulation and the resulting coupled governing equations for the present model. The temporal evolution of the spherical and shape mode oscillations are discussed in section~\ref{sec:temp}. Section~\ref{sec:per} provides the analysis of applying the Krylov-Bogoliubov perturbation technique to the governing equation. This is followed by the steady-state solutions of the slow-time equations and the conditional stability analysis in section~\ref{sec:stab}. The conclusions and an outlook of further work are provided in section~\ref{sec:con}.

\section{Kinematics of deformation}
\label{sec:kindef}

\begin{figure}[t!]
    \centering
    \includegraphics[width=0.6\linewidth]{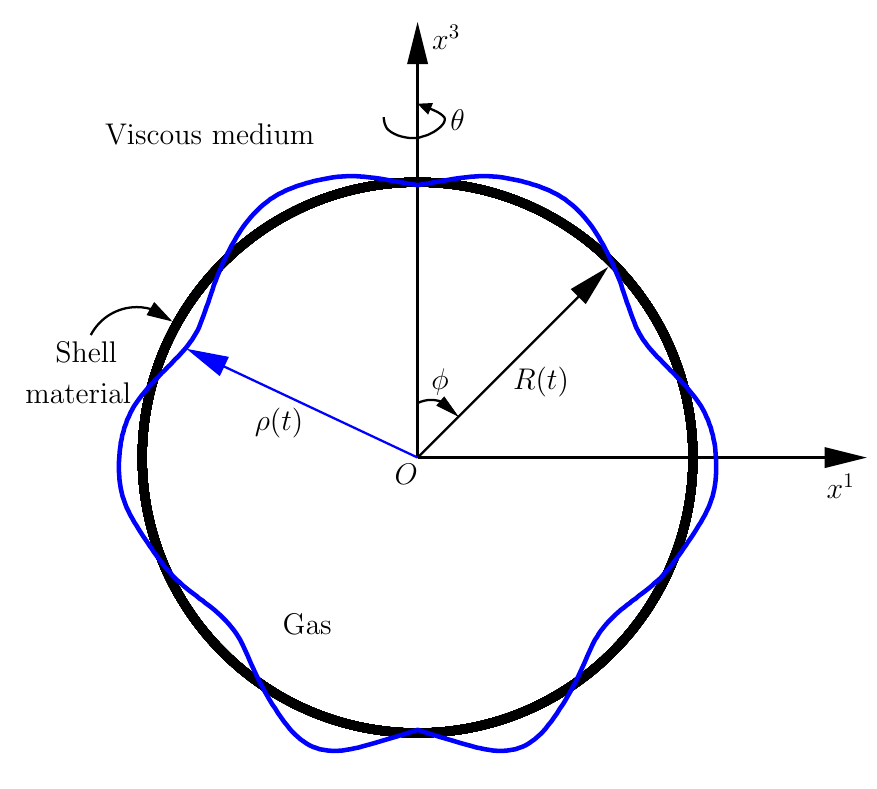}
    \caption{The deformed configuration of a thin encapsulated microbubble filled with gas and suspended in viscous medium.}
    \label{f:1}
\end{figure}

Consider a gas-filled encapsulated spherical bubble suspended in an infinite incompressible viscous medium. In the present analysis, the encapsulation is treated as a thin mathematical surface, and the gas-encapsulation and encapsulation-liquid interfaces are assumed to have zero thickness, following the nonlinear continuum framework of \citet{steigmann}. These interfaces are also referred to as \textit{Steigmann-Ogden interfaces} (SOIs). \citep{dash_tamadapu_2022} The equilibrium radius and the encapsulation thickness of the bubble are denoted by $R_{0}$ and $q$, respectively. A schematic of the bubble's deformed geometry suspended in a viscous medium with the frame of reference is shown in figure~\ref{f:1}.

\subsection{Reference configuration}
When no acoustic fields are applied, the bubble is considered to be in its reference configuration $\mathfrak{D}_0$.  
Assume that $\{X^{\rm1},X^{\rm2},X{\rm^3}\}$ be the three-dimensional Cartesian coordinate system with unit basis vector triad as $\{\bs{e}_1,\bs{e}_2,\bs{e}_3\}$, and that $\{r,\phi,\theta\}$ be the principal spherical polar coordinates with origin at the center and unit basis vector triad $\{\bs{e}_r,\bs{e}_\phi,\bs{e}_\theta\}$. Here, $\{r,\phi,\theta\}$ corresponds to the radial distance from the centre of the bubble, polar angle formed with $X^{3}-$axis, and azimuthal angle measured about $X^3-$axis in the counter clockwise direction from $X^1-$axis, respectively. The coordinates of a material particle at a point in the undeformed reference configuration is given by the following relation between the Cartesian and the spherical polar coordinates as
\begin{align}
\begin{split}
    X^{1}&=r\sin\phi\cos\theta,\\
    X^{2}&=r\sin\phi\sin\theta,\\
    X^{3}&=r\cos\phi.
\end{split}
\end{align}

\subsection{Deformed configuration}
Under the action of acoustic field, the bubble deforms and attains a deformed configuration $\mathfrak{D}$. The coordinates of the same typical point at a given time instant $t$ in an arbitrary deformed configuration, with the Cartesian coordinates $\{x^{\rm1},x^{\rm2},x{\rm^3}\}$ and the spherical polar coordinates $\{\rho(\phi,t),\phi,\theta\}$ is given by
\begin{align}
\begin{split}
    x^{1}&=\rho(\phi,t)\sin\phi\cos\theta,\\
    x^{2}&=\rho(\phi,t)\sin\phi\sin\theta,\\
    x^{3}&=\rho(\phi,t)\cos\phi.
\end{split}
\end{align}

\subsection{Interface energy model}
\citet{steigmann} proposed a nonlinear continuum framework to describe the kinematics of the elastic surface-substrate interactions. Later, \citet{GAO201459} developed a more general interface theory considering the effects of curvature-dependent interface energy and the residual elastic field in the bulk induced by this interface energy. Following these works, \citet{dash2022describing} and \citet{dash_tamadapu_2022} proposed an interface energy model within the framework of surface continuum mechanics to study the radial dynamics of an encapsulated microbubble. In this section, we will provide some preliminary notations and definitions to describe the kinematics of deformation of these interfaces. For detailed explanations and derivations the reader may refer to\citet{steigmann,GAO201459,dash2022describing,dash_tamadapu_2022}. 

Let $\theta^1=\phi$ and $\theta^2=\theta$ be the surface coordinates of the spherical interface with radius $r$. Consider $\bs{Z}(\theta^1,\theta^2)$ and $\bs{z}(\theta^1,\theta^2)$ as the position vectors of the same point on the undeformed and deformed interfaces. Assuming that the interface is convected by the deformation of the bulk of the bubble, the same material point before and after the deformation can be related using the deformation mapping $\bs{\chi}$ such that
\begin{align}
    \bs{z}(\theta^1,\theta^2)=\bs{\chi}\left(\bs{Z}(\theta^1,\theta^2)\right).
\end{align}
The respective tangent vectors $\bs{G}_\alpha$ and $\bs{g}_\alpha$ on the undeformed and deformed interfaces induced by these coordinates, are given by
\begin{align}
\begin{split}
    \bs{G}_\alpha=\bs{Z}_{,\alpha},\quad
    \bs{g}_\alpha=\bs{z}_{,\alpha},\quad
    \alpha\in\{1,2\},\quad
    (\cdot)_{,\alpha}=\f{\del(\cdot)}{\del\theta^\alpha}.
\label{e:tgtvec}
\end{split}
\end{align}
The components of the covariant metric tensor in the undeformed and deformed interfaces, respectively, are given by
\begin{align}
    G_{\alpha\beta}&=\text{diag}\left[r^2,r^2 \sin^2{\phi}\right], \\
    g_{\alpha\beta}&=\text{diag}\left[{\rho(\phi,t)}^2+{\rho'(\phi,t)}^2,{\rho(\phi,t)}^2\sin^2\phi\right],
\end{align}
where the prime ${(\cdot)}'$ denotes the derivative with respect to $\phi$.
The components of the mixed right Cauchy-Green deformation tensor can be obtained as
\begin{align}
    {C^\alpha}_\beta=G^{\alpha\delta}\,g_{\delta\beta}=\text{diag}\left[\f{{\rho(\phi,t)}^2+{\rho'(\phi,t)}^2}{r^2},\f{{\rho(\phi,t)}^2}{r^2}\right].
\end{align}
The second fundamental forms which represent the normal curvatures for the undeformed and deformed interfaces are denoted by $Q_{\alpha\beta}$ and $q_{\alpha\beta}$, respectively, given by
\begin{align}
    Q_{\alpha\beta}=\bs{N}\cdot\bs{G}_{\alpha,\beta},\quad 
    q_{\alpha\beta}=\bs{n}\cdot\bs{g}_{\alpha,\beta}.
\label{e:qtensor}
\end{align}
where $\bs{N}$ and $\bs{n}$ are the oriented unit normals to the undeformed and deformed interfaces, respectively. By definition, the relative curvature tensor $\kappa_{\alpha\beta}=-q_{\alpha\beta}$ and its co-variant components are given by
\begin{align}
    \kappa_{\alpha\beta}=\text{diag}\left[\kappa_{11},\kappa_{22}\right],
\end{align}
where 
\begin{align}
    \kappa_{11}&=-\f{{\rho(\phi,r)}^2+2{\rho'(\phi,t)}^2-\rho(\phi,t)\rho''(\phi,t)}{\sqrt{{\rho(\phi,t)}^2+{\rho'(\phi,t)}^2}},\nn\\
    \kappa_{22}&=-\f{\rho(\phi,t)\sin(\phi)\left[\rho(\phi,t)\sin{\phi}-\rho'(\phi,t)\cos{\phi}\right]}{\sqrt{{\rho(\phi,t)}^2+{\rho'(\phi,t)}^2}}\nn.
\end{align}
The contra-variant components of the relative curvature tensor $(\kappa^{\alpha\beta})$ are obtained using the contraction operation
\begin{align}
    \kappa^{\alpha\beta}=G^{\alpha\gamma}G^{\beta\delta}\kappa_{\gamma\delta}.
\end{align}
The adjugate $(\td \cdot)$ of the symmetric right Cauchy-Green deformation tensor $\td C^{\alpha\beta}$, and the relative curvature tensor $\td \kappa^{\alpha\beta}$ are defined by \citet{steigmann}
\begin{align}
    \td C^{\alpha\beta}&=\left(g/G\right)g^{\alpha\beta},\\
    \td \kappa^{\alpha\beta}&=\mu^{\alpha\beta}\mu^{\beta\lambda}\kappa_{\gamma\lambda},  
\end{align}
where $g={\rm{det}}\left(g_{\alpha\beta}\right)$, $G={\rm{det}}\left(G_{\alpha\beta}\right)$, $\mu^{\alpha\beta}=e^{\alpha\beta}/\sqrt{G}$, and $e^{\alpha\beta}=e_{\alpha\beta}$ is the alternator symbol.

The gas-encapsulation and the encapsulation-liquid interfaces are considered as hemitropic interfaces, and the energy density can be expressed as a function of right Cauchy-Green interface deformation tensor $\msb{C}$ and the relative curvature tensor $\bs{\kappa}$ \citep{yuan-chengfung1977,steigmann,GAO201459,dash_tamadapu_2022} as
\begin{align}
    \gamma=\gamma(\msb{C},{\bs \kappa}),
\end{align}
and satisfies the relation
\begin{align}
    \gamma(\msb{C},{\bs \kappa})=\gamma(\msb{Q}\msb{C}\msb{Q}^{\rm T},\msb{Q}{\bs \kappa}\msb{Q}^{\rm T}),
\end{align}
where $\msb{Q}$ is a proper-orthogonal second-order tensor.
The Cauchy interface stress $(\bs\sigma)$ and the moment $(\msb m)$ tensors are calculated using the relations
\begin{align}
    J {\bs \sigma}=2\,\f{\del \gamma}{\del\msb C},\quad {\msb m}=\f{\del\gamma}{\del \bs\kappa}.
\end{align}
Further, the expressions for the components $(T^{\alpha\beta})$ of the interface stress tensor $\left(\msb{T}\right)$ and the components $(M^{\alpha\beta})$ of the bending moment tensor $\left(\msb{M}\right)$ are obtained in the form of the following constitutive equations (for detailed derivation see \citet{steigmann,GAO201459,dash_tamadapu_2022})
\begin{align}
\begin{split}
    T^{\alpha\beta}&=\f{1}{2}J\sigma^{\alpha\beta}=\f{\del \gamma}{\del I_1}G^{\alpha\beta}+
    \f{\del \gamma}{\del I_2}\td{C}^{\alpha\beta}+\f{\del \gamma}{\del I_5}\kappa^{\alpha\beta}+\f{1}{2}\f{\del \gamma}{\del I_6}\left(D^{\alpha\beta}+D^{\beta\alpha}\right),\\
    M^{\alpha\beta}&=Jm^{\alpha\beta}=\f{\del \gamma}{\del I_3}G^{\alpha\beta}+
    \f{\del \gamma}{\del I_4}\td{\kappa}^{\alpha\beta}+\f{\del \gamma}{\del I_5}C^{\alpha\beta}+\f{1}{2}\f{\del \gamma}{\del I_6}\left(E^{\alpha\beta}+E^{\beta\alpha}\right),
    \label{e:istr}
\end{split}
\end{align}
where
\begin{align}
    J=\sqrt{g/G},\quad D^{\alpha\beta}=G_{\gamma\delta}\mu^{\alpha\gamma}\kappa^{\beta\delta},\quad
    E^{\alpha\beta}=G_{\gamma\delta}\mu^{\alpha\gamma}C^{\beta\delta},
\end{align}
and $(I_1, I_2, I_3, I_4, I_5, I_6)$ are the six basis invariants (defined in appendix~\ref{app:A0}) of the right Cauchy-Green interface deformation tensor $\msb{C}$, the relative curvature tensor $\bs{\kappa}$, and the permutation tensor-density $\bs{\mu}$ on the undeformed interface.

\section{Mathematical model}
\label{sec:mathmodel}

The following section presents the mathematical model formulation for the present problem. Section \ref{ssec:edf} outlines the energy density functions for the shell material. The first variational formulations and the resulting governing equations are discussed in subsections \ref{ssec:varf} and \ref{ssec:ge}, respectively. 

\subsection{Material energy density function for the thin shell membrane}
\label{ssec:edf}
The shell material and surrounding fluid are considered to be homogeneous, isotropic and incompressible. The bubble shell material assumed to be a hyperelastic membrane following the constitutive relation of the neo-Hookean material model. The neo-Hookean material is chosen because it offers a simple and realistic model for a rubbber-elastic type material. For the incompressible bulk neo-Hookean material model, the strain energy density function (per unit undeformed volume) is given by
\begin{align}
    \Psi_{\rm s}=C_1(\td I_1-3),
\end{align}
where $C_1$ is the material elastic constant related to the shear modulus $\mu=2\,C_1$, and $\td I_1$ is the first invariant of the right-Cauchy Green deformation tensor $\msb C$. Thus, $\td I_1$ is given by
\begin{align}
    \td I_1=\rm{tr}\,\msb{C}=\Lambda_1^2+\Lambda_2^2+\Lambda_3^2,
\end{align}
where $(\Lambda_1,\Lambda_2,\Lambda_3)$ are the principal stretch ratios in the radial (across the thickness), meridional, and azimuthal directions, respectively. The expressions for the principal stretch ratios are given by
\begin{align}
    \Lambda_1=\f{r^2}{\rho(\phi,t)\sqrt{{\rho(\phi,t)}^2+{\rho'(\phi,t)}^2}},\quad \Lambda_2=\f{\sqrt{{\rho(\phi,t)}^2+{\rho'(\phi,t)}^2}}{r},\quad \Lambda_3=\f{{\rho(\phi,t)}}{r}.
\end{align}
Here the stretch $(\Lambda_1)$ across the radial (thickness) direction has been calculated using the incompressibility constraint $\Lambda_1\Lambda_2\Lambda_3=1$. 

\subsection{Variational formulation}
\label{ssec:varf}

In addition to those in \citet{guedra2018bubble}, the new terms in the governing equations related to the interface energy and viscoelastic shell are obtained using a variational formulation of the energy functional. The energy functional $(\Pi)$ for the present problem constitutes of three parts: the interface energy $(\Pi_{\rm{ie}})$, and energy of the viscoelastic shell $(\Pi_{\rm{s}})$, such that
\begin{align}
    \Pi=\Pi_{\rm{ie}}+\Pi_{\rm{s}},
\end{align}
and the expressions for the respective energy functional are given by 
\begin{align}
    \Pi_{\rm{ie}}=\int_0^{\pi} \gamma\left(\msb{C},\bs \kappa\right) \sqrt{G}\,\text{d}\phi,\quad
    \Pi_{\rm{s}}=\int_0^{\pi} \Psi_{\rm{s}}\;q\sqrt{G}\,\text{d}\phi,
\end{align}
where $\sqrt{g}=r^2 \sin \phi$. 
Using the variational principle, the total energy functional is minimized $(\delta\Pi=0)$ to obtain respective terms in the governing equilibrium equations. Since the interface energy density is a function of $\msb{C}$ and $\bs{\kappa}$, its variation can further be written as \citep{GAO201459}
\begin{align}
    \delta\left(\gamma\left(\msb{C},\bs \kappa\right)\right)=\f{\del \gamma}{\del \msb{C}}:\delta \msb{C}+\f{\del \gamma}{\del \bs{\kappa}}:\delta \bs{\kappa}=\msb{T}:\delta \msb{C}+\msb{M}:\delta\bs{\kappa}.
\end{align}
Therefore, the first variation of the energy functional is given by
\begin{align}
    \delta\Pi=\int_0^{\pi} \left(\msb{T}:\delta \msb{C}+\msb{M}:\delta\bs{\kappa}\right)\sqrt{G}\,\text{d}\phi+\int_0^{\pi} \delta\left(\Psi_{\rm{s}}\right) q\sqrt{G}\,\text{d}\phi.
\label{e:vareq}
\end{align}
The expression for $\delta \msb{C}$ and $\delta\msb{\kappa}$ can be further expressed in terms of the $\delta R$ and $\delta a_n$ to obtain additional terms in the governing differential equation.  

\subsection{Equations governing the spherical mode and the shape mode}
\label{ssec:ge}
In this study, the wavelength of the acoustic field is considered to be sufficiently large such that it can be assumed to behave uniformly on the bubble surface. For simplicity, we neglect the translational motion of the bubble. Assuming that the motion of the spherical bubble is dominated in the radial direction, the perturbation to the spherical bubble can be expanded in terms of spherical harmonics. Since we restrict attention to small axisymmetric shape deformation of the bubble, the spherical harmonics reduces to Legendre polynomials. At time $t$, the surface of the bubble $\rho\left(\phi,t\right)$ is given by
\begin{align}
    \rho\left(\phi,t\right)=R\left(t\right)+\sum_{n=2}^{\infty} \epsilon a_n(t) P_n\left(\cos{\phi}\right),
\label{e:bubbsur}
\end{align}
where $R(t)$ is the radius of the spherical bubble (spherical or volume mode), $P_n(\cos \phi)$ represents the Legendre polynomial of order $n$, and $\epsilon\,a_n(t)$ is the amplitude of the $n$th Legendre mode, commonly referred to as shape mode amplitude. Here, the surface/shape distortion terms of the bubble radius $R(t)$ are assumed to be small, therefore their amplitudes $a_n(t)$ are scaled by the small parameter $\epsilon$, whereas no restrictions were imposed on the spherical/volume oscillations.

\citet{Shaw2006} developed the Lagrangian formulation for a gas bubble in an incompressible liquid of infinite extent. By integrating the Lagrangian density across the problem domain, the resulting Lagrangian $\mathcal{L}$ is obtained and can be written as
\begin{align}
    \mathcal{L}=\mathcal{C}-\mathcal{T}-\mathcal{K}.
\end{align}
where $\left(\mathcal{C},\mathcal{T},\mathcal{K}\right)$ stands for the constraint, potential energy and kinetic energy terms, respectively. Their respective expressions are given by \citet{Shaw2006}
\begin{align}
    \mathcal{C}&=-2\pi\rho^{\rm L} \int_0^{2\pi} \Phi\left(\rho \f{\del \rho}{\del t}\right)\rho \sin\phi\,\rm{d}\phi,\\
    \mathcal{T}&=\sigma_{(\cdot)}\,\rvert S \rvert+\Pi-V\left[\f{p_{g_0}}{1-k}{\left(\f{V_0}{V}\right)}^k-p_\infty\right],\\
    \mathcal{K}&=-\pi\rho^{\rm L}\int_{-1}^{1}\Phi f_1\, \rm{d}(\cos\phi),
\end{align}
where
\begin{align}
    \rvert S\rvert&=2\pi\int_0^{\pi}\rho^2\left[1+\f{1}{\rho^2}{\left(\f{\del \rho}{\del \phi}\right)}^2\right]^{1/2}\,\sin\phi\,\rm{d}\phi,\\
    f_1&=b_0(t)-\sum_{n=1}^{\infty}\f{\epsilon b_n(t)(n+1)}{\rho^n}P_n(\cos\phi)-(1-\cos^2\phi)\nn\\
    &\qquad \times\left(\sum_{n=2}^\infty\epsilon a_n(t)\f{\text{d}P_n}{\text{d}(\cos\phi)}\right)\left(\sum_{n=2}^\infty \f{\epsilon b_n(t)}{\rho^{n+1}}\f{\text{d}P_n}{\text{d}(\cos\phi)}\right),
\end{align}
$\rho^{\rm L}$ is the density of liquid, $V$ is the volume at any instant of time $t$, $V_0$ is the initial volume, $\Phi$ is obtained by evaluating $\varphi$, defined in \eqref{e:velpot}, on the bubble surface, and $\sigma_{(\cdot)}$ is the respective effective surface tension parameter for spherical and nonspherical modes. 

Consistent with much of other works, they defined the bubble surface as in \eqref{e:bubbsur} and the general fluid velocity potential $\varphi$ as
\begin{align}
    \varphi=-\f{b_0(t)}{r}+\sum_{n=1}^\infty \epsilon \f{b_n(t)}{r^{n+1}}P_n(\cos\phi).
    \label{e:velpot}
\end{align}
The expansions for $\rho\left(\phi,t\right)$ and potential $\varphi$ evaluated on the bubble surface are substituted into the respective expressions of constraint, potential, and kinetic energy. The resulting expressions are truncated to $\mathcal{O}(\epsilon^3)$ to obtain the expression for Lagrangian $\mathcal{L}$ in terms of parameters $R(t),\dot{R}(t),a_n(t), \dot{a}_n(t),$ and $b_n(t)$ to order $\epsilon^3$. The coefficients of $b_n(t)$ are then eliminated from the Lagrangian $\mathcal{L}$ by applying the kinematic boundary conditions imposing the set of following conditions
\begin{align}
    \f{\del \mathcal{L}}{\del b_{n}}=0, \qquad n=0,1,2,...
\end{align}
From the above conditions, we get 
\begin{align}
    b_0(t)=R^2\dot{R},\qquad
b_n(t)=-\f{R^{n+1}}{n+1}\left(2a_n\dot{R}+\dot{a}_n R\right)+\mathcal{O}(\epsilon).
\end{align}
The resultant Lagrangian is then used in the Euler-Lagrange equations to obtain the equations governing the spherical, and shape oscillations. A detailed description of the Lagrangian formulation can be found in \citet{Shaw2006}.
By following the mathematical formulation presented in \citet{Shaw2006}, and using the orthogonality of Legendre polynomials, the integrals are deduced further to obtain additional terms in the governing equations. Incorporating the effects of interface energy (through the interface tension), and viscosity and elasticity of the shell membrane, the present model introduces additional terms in \eqref{e:smode} and \eqref{e:nsmode} beyond those considered by \citet{guedra2018bubble}. Thus, the resulting equations governing the spherical mode $R(t)$ and the nonspherical (shape) modes $a_n(t)$ accurate to second order approximation in $\epsilon$ with $n\geqslant2$ are given, respectively, by
\begin{align}
&\ddot R+\f{3}{2}\f{\dot R^2}{R}-\f{1}{\rho^{\rm L}
R}\left[p_{g_0}{\left(\f{R_0}{R}\right)}^{3k}-p_\infty\right]+\f{2\sigma_{\rm sp}}{\rho^{\rm L} R^2}
+4\nu^{\rm S}\f{\dot R}{R^2}+\f{4 C_1 q}{\rho^{\rm L} R^2}\left[1-\f{R_0^6}{R^6}\right]\nn\\
&+\f{\epsilon^2}{R}\sum_{n=2}^{\infty}\Bigg\{\f{1}{\mathcal{H}_n}\left[\left(n+\f{3}{2}\right)\dot a_n^2+\left(n+3\right)a_n\ddot{a}_n-\left(n-3\right)\left(\f{\ddot R}{R}a_n^2+\f{1}{2}\f{\dot R^2}{R^2}\,a_n^2+2\f{\dot R}{R}a_n \dot a_n\right)\right]\Bigg\} \nn\\
&+ \f{\epsilon^2}{\rho R^3}\left[p_\infty-p_{g_0}\left(1-3k\right)\left(\f{R_0}{R}\right)^{3k}\right]\sum_{n=2}^{\infty}\,\f{a_n^2}{\left(2n+1\right)}+\epsilon^2\f{6 C_1 q R_0^6}{\rho^{\rm L} R^{10}}\left[\sum_{n=2}^{\infty}\,\f{(n(n+1)-10)}{2n+1}a_n^2\right]=0,
\label{e:smode}
\end{align}
and
\begin{align}
&\epsilon\ddot a_n+\epsilon\left(\f{3\dot R}{R}+F_\nu \right)\dot a_n+\epsilon(n+1)\f{\sigma_{\rm nsp}^{0}}{\rho^{\rm L} R^3}a_n+\epsilon\left(n-1\right)\left(G_\nu-\f{\ddot R}{R}\right)a_n\nn\\
&+\epsilon\,(n+1)a_n\f{2C_1 q}{\rho^{\rm L} R^3}\left[\big\{(n-1)(n+2)\big\}+\big\{14-n(n+1)\big\}\f{R_0^6}{R^6}\right]+\epsilon^2\f{\mathcal{H}_n}{2 \rho^{\rm L} R^4}\nn\\
&\times\left[\sum_{i=2}^\infty\sum_{j=2}^\infty \sigma_{\rm nsp}^{1} a_i a_j\right] 
+\epsilon^2\f{\mathcal{H}_n}{4R}\sum_{i=2}^{\infty}\,\sum_{j=2}^{\infty}\Bigg\{\f{\ddot R}{R}a_i a_j G_{d_{ijn}}+\f{\dot R^2}{R^2}a_i a_j M_{a_{nij}} +\f{\dot R}{R} a_i \dot a_j M_{b_{nij}}\nn\\
&+a_i \ddot a_j M_{c_{nij}}
+\dot a_i \dot a_j M_{d_{nij}}\Bigg\}+\epsilon^2\f{\mathcal{H}_n}{2\rho^{\rm L} R^3}\left[p_\infty-p_{g_0}\left(\f{R_0}{R}\right)^{3k}\right]\sum_{i=2}^{\infty}\sum_{j=2}^{\infty}a_i a_j I_{a_{nij}}\nn\\
&-\epsilon^2\f{6 C_1 q \mathcal{H}_n R_0^6}{\rho^{\rm L} R^{10}}
\sum_{i=2}^\infty\sum_{j=2}^\infty\left(5 I_{a_{nij}}+\f{1}{2}I_{c_{nij}}+ I_{c_{ijn}}\right)a_i a_j=0.
\label{e:nsmode}
\end{align}
Here, $\mathcal{H}_n=(2n+1)(n+1)$, $R_0$ is the radius of bubble at static equilibrium configuration, $\nu^{\rm L}$ and $\nu^{\rm S}$ denote the kinematic viscosity of surrounding liquid and bubble shell material, respectively, and $k$ is the polytropic expansion index. The gas pressure inside the bubble $\left(p_{g_0}\right)$ is represented in terms of static liquid pressure $(p_0)$ and the surface tension parameter at static equilibrium $(\sigma_{\rm eq})$ given by
\begin{align}
    p_{g_0}=p_0+\f{2\sigma_{\rm eq}}{R_{0}},
\end{align}
with
\begin{align}
    \sigma_{\rm eq}&=\gamma_0+2\gamma_1+2\gamma_2+\gamma_3\f{1}{R_0}
    +\gamma_4\f{1}{R_0^2}+3\gamma_5\f{1}{R_0},
\end{align}
where $\gamma_k=\gamma_{1k}+\gamma_{2k}$ for $k=\{0,1,2,4\}$ and $\gamma_l=-\gamma_{1l}+\gamma_{2l}$ for $l=\{3,5\}$ are the interface constants with $\gamma_{ij},i=1,2,\,j=1\,\text{to}\,5$ as interface parameters at the gas--encapsulation and encapsulation--liquid interfaces, respectively. In \citet{dash_tamadapu_2022} and \citet{dash2022describing}, the interface parameters for the gas-encapsulation and encapsulation-liquid interfaces are associated with the inner and outer radii of the bubble, respectively. In the present work, mathematical surface treatment of the bubble oscillations in terms of single radial parameter $R(t)$ leads to the net interface parameters $\gamma_i$ in governing equations. 
It is important to note that the orientation of the normal vector plays an important role for the interface parameters connected to the curvature tensor.
\citep{dash_tamadapu_2022,dash2022describing}

The liquid pressure $p_\infty\left(t\right)$ at any instant of time $t$ is 
\begin{align}
    p_\infty\left(t\right)=p_0-p_a\cos\left({2\pi ft}\right),
\end{align}
where $p_a$ is acoustic pressure excitation. The terms $(\sigma_{\rm sp},\sigma_{\rm nsp}^0,\sigma_{\rm nsp}^1)$ represent the effective surface tension in terms of interface parameters $\left(\gamma_{ij}\right)$ for the spherical $(\cdot)_{\rm sp}$ and nonspherical $(\cdot)_{\rm nsp}$ modes, respectively, expressed as follows
\allowdisplaybreaks{
\begin{subequations}
    \begin{align}
\sigma_{\rm sp}&=\gamma_{0}+2\left[\gamma_1+\gamma_2\f{R^2}{R_0^2}+\gamma_5\f{R}{R_0^2}\right]+\f{1}{R}\bigg[\gamma_3+\gamma_4\f{R}{R_0^2}+\gamma_5\f{R^2}{R_0^2}\bigg]\nn\\
&+\f{\epsilon^2}{R^2}\sum_{n=2}^{\infty}\f{a_n^2}{2n+1}\bigg[\big\{n(n+1)+6\big\}\gamma_2\f{R^2}{R_0^2}-\f{n(n+1)}{2R}\gamma_3+\big\{2n(n+1)+3\big\}\gamma_5\f{R}{R_0^2}\bigg], \label{e:sig1sm}\\
\sigma^0_{\rm nsp}&=(n-1)(n+2)\Bigg[\gamma_0+ 2\gamma_1+ 2\gamma_2\f{R^2}{R_0^2}+2\gamma_3\f{1}{R}+5\gamma_4\f{1}{R_0^2}+8\gamma_5\f{R}{R_0^2}\Bigg]\nn\\
&+4\gamma_2\f{R^2}{R_0^2}+8\gamma_4\f{1}{R_0^2}+16\gamma_5\f{R}{R_0^2}, \label{e:sig2nsm} \\
\sigma_{\rm nsp}^1&=N^2_{nij}\gamma_2\f{ R^2}{R_0^2}+N^3_{nij}\gamma_3\f{1}{R}+N^4_{nij}\gamma_4\f{1}{R_0^2}+N^5_{nij}\gamma_5\f{R}{R_0^2}.\label{e:sig3nsm}
\end{align}
\label{e:sig}
\end{subequations}}

In the present study, the viscous dissipation in the surrounding fluid and the shell membrane has been introduced in a \textit{ad hoc} manner into equations \eqref{e:smode} and \eqref{e:nsmode}. Following the discussions presented in \citet{guedra2018bubble}, these terms are introduced as classical viscous terms to lower orders. In the equation governing the spherical mode of oscillation, the influence of viscosity arising from both the bubble shell and the surrounding fluid is introduced using the classical term, similar to the equation governing the radial oscillations of the bubble. The viscous effects associated with the surrounding fluid in the shape mode equation are incorporated using established functions within the boundary layer approximation such that
\begin{subequations}
\begin{align}
    F_\nu&=(n+2)\left[(2n+1)\nu^{\rm S}-2n(n+2)\f{\delta}{R}\nu^{\rm L}\right]\f{2}{R^2},\\
    G_\nu&=(n+2)\left[\nu^{\rm S}+2n\f{\delta}{R}\nu^{\rm L}\right]\f{2}{R^2}\f{\dot R}{R},
\end{align}
\end{subequations}
where $\delta=\sqrt{\nu^{\rm L}/\omega}$ represents the viscous boundary layer thickness. \citep{brenner1995bubble} A comprehensive mathematical analysis has been conducted by \citet{shaw2009stability}, taking into account the combined impact of the Rayleigh dissipation function and perturbation analysis, in order to address the effects of viscosity. This would result in nonlinear viscous terms at higher orders in \eqref{e:smode} and \eqref{e:nsmode}. However, \citet{guedra2018bubble} considered fluid viscosity as a small perturbation, resulting in the neglect of such higher-order terms in the subsequent asymptotic expansion of both the spherical and nonspherical equations anyway. Following a similar approach, we consider that both the fluid and shell viscosity act as small perturbations. This simplifies the analysis and drops the nonlinear viscous terms appearing at the higher-orders expansions. The set of integrals (see appendix~\ref{app:A}), and the other coefficients appearing the subsequent calculations are listed in appendix~\ref{app:B}.

\begin{table}
\caption{\label{tb:1} The numerical values of inner and outer interface parameters such as $(\gamma_{11},\gamma_{21},\gamma_{12},\gamma_{22})\,$N/m, $(\gamma_{14},\gamma_{24})\,$Nm and $(\gamma_{13},\gamma_{23},\gamma_{15},\gamma_{25})\,$N of the bubble chosen for numerical simulations.}
\begin{ruledtabular}
\begin{tabular}{cccccccccc}
$\gamma_{11}$ & $\gamma_{21}$ & $\gamma_{12}$ & $\gamma_{22}$ & $\gamma_{13}$ & $\gamma_{23}$ & $\gamma_{14}$ & $\gamma_{24}$ & $\gamma_{15}$ & $\gamma_{25}$ \\ \hline
0.01 & 0.01 & 0.01 & 0.01 & 0.04 & 0.035 & 0.03 & 0.01 & 0.04 & 0.03
\end{tabular}
\end{ruledtabular}
\end{table}

The primary focus of this work is to understand the nonspherical oscillations of smaller radii microbubbles, where the influence of interface parameters becomes particularly important. Therefore, we consider the equilibrium radius of the bubble $R_{0}=2\,\mu$m with shell membrane thickness of $q=20\,$nm. The density of the surrounding liquid is considered to be $\rho^{\rm L}=1000\,$kg/${\rm{m^3}}$, viscosity of the fluid $(\nu^{\rm L})$ and shell $(\nu^{\rm S})$ are considered to be $(10^{-6})\,\rm{m^2/s}$, the shell material elastic constant $C_1=0.1\,$MPa, and static liquid pressure $p_0=0.1\,$MPa. The values of these physical and material parameters remain consistent throughout the numerical simulations, unless otherwise specified.
Given our interest in studying an EB with a radius of $\mathcal{O}\left(10^{-6}\right)$m and a thickness of $\mathcal{O}\left(10^{-9}\right)$m, we assume the order of interface parameters $\gamma_{10},\gamma_{20},\gamma_{11},\gamma_{21},\gamma_{12},\gamma_{22}$ as $\mathcal{O}(1)\,$N/m, $\gamma_{13},\gamma_{23},\gamma_{15},\gamma_{25}$ as $\mathcal{O}(10^{-6})\,$N and $\gamma_{14},\gamma_{24}$ as $\mathcal{O}(10^{-12})\,$N\,m. The order of these interface parameters are such that the effective interface tension parameters possesses reasonable values of $\mathcal{O}(1) $.\citep{dash2022describing,dash_tamadapu_2022} The values of these interface parameters for the present study are tabulated in table~\ref{tb:1}. The coupled governing equations \eqref{e:smode} and \eqref{e:nsmode} indicate the complex interactions between spherical and shape deformation of an EB. It is also apparent that numerical studies using such mathematical model may offer many possibilities that can be explored. Nevertheless, the present model introduces the interface parameters as a preliminary, yet comprehensive, study that highlights their substantial influence on the oscillations of smaller radii bubbles.

\section{Temporal evolution of spherical and shape mode amplitudes}
\label{sec:temp}

\begin{figure}
    \centering
    \includegraphics[width=\linewidth]{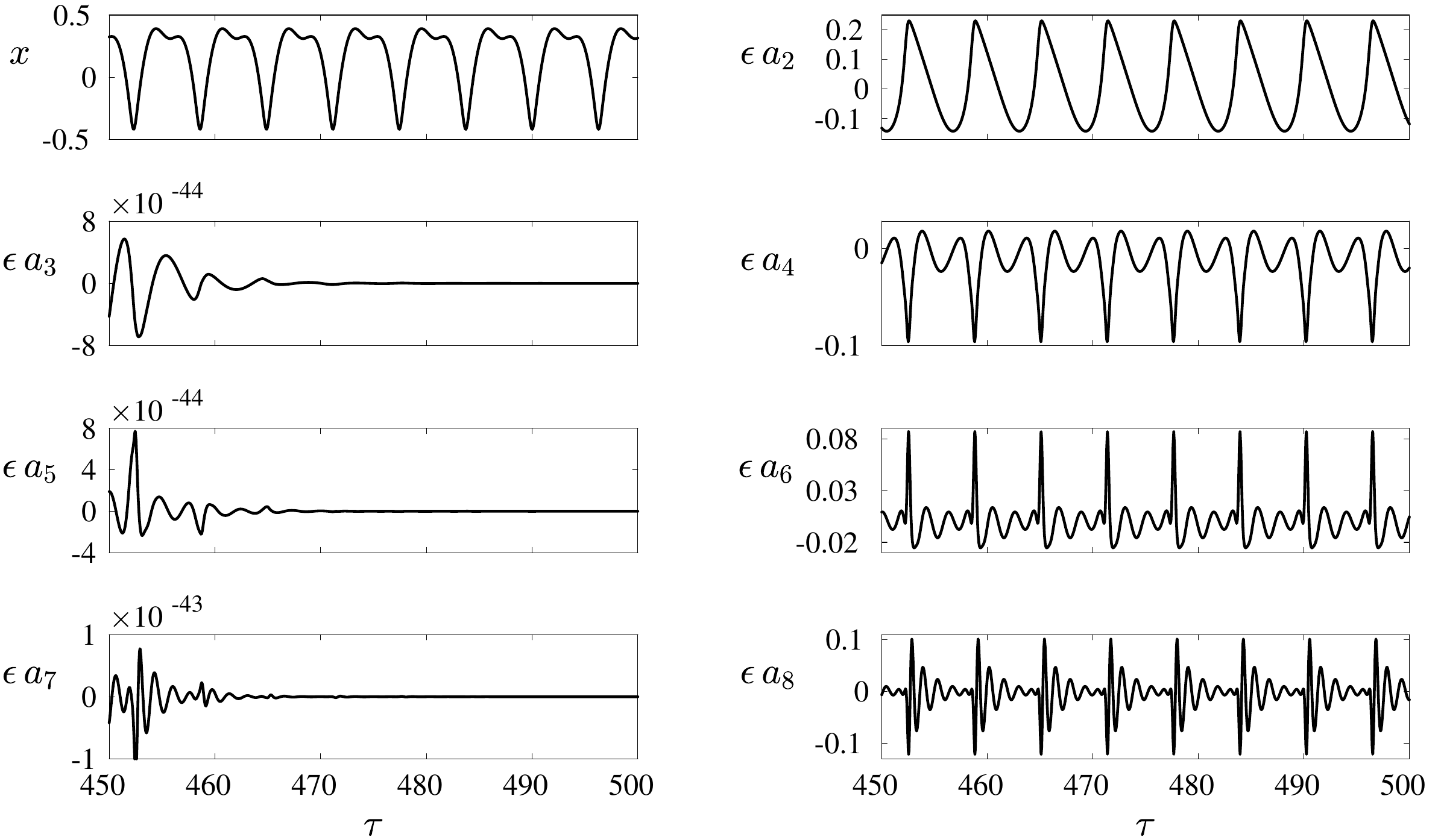}
    \caption{Steady state oscillations of the spherical and the seven first shape modes for the EB with $R_0=2\,\mu$m driven above the first parametric resonance of the $n=2$ mode with $p_a=0.37\,$MPa, $f=4.96\,$MHz, and $\Omega_{0,n}\sim1$ (corresponding to the point $\bullet$ in figure~\ref{f:n2stab}) with the set of interface parameters in table~\ref{tb:1}.}
    \label{f:n2}
\end{figure}

\begin{figure}
    \centering
    \includegraphics[width=\linewidth]{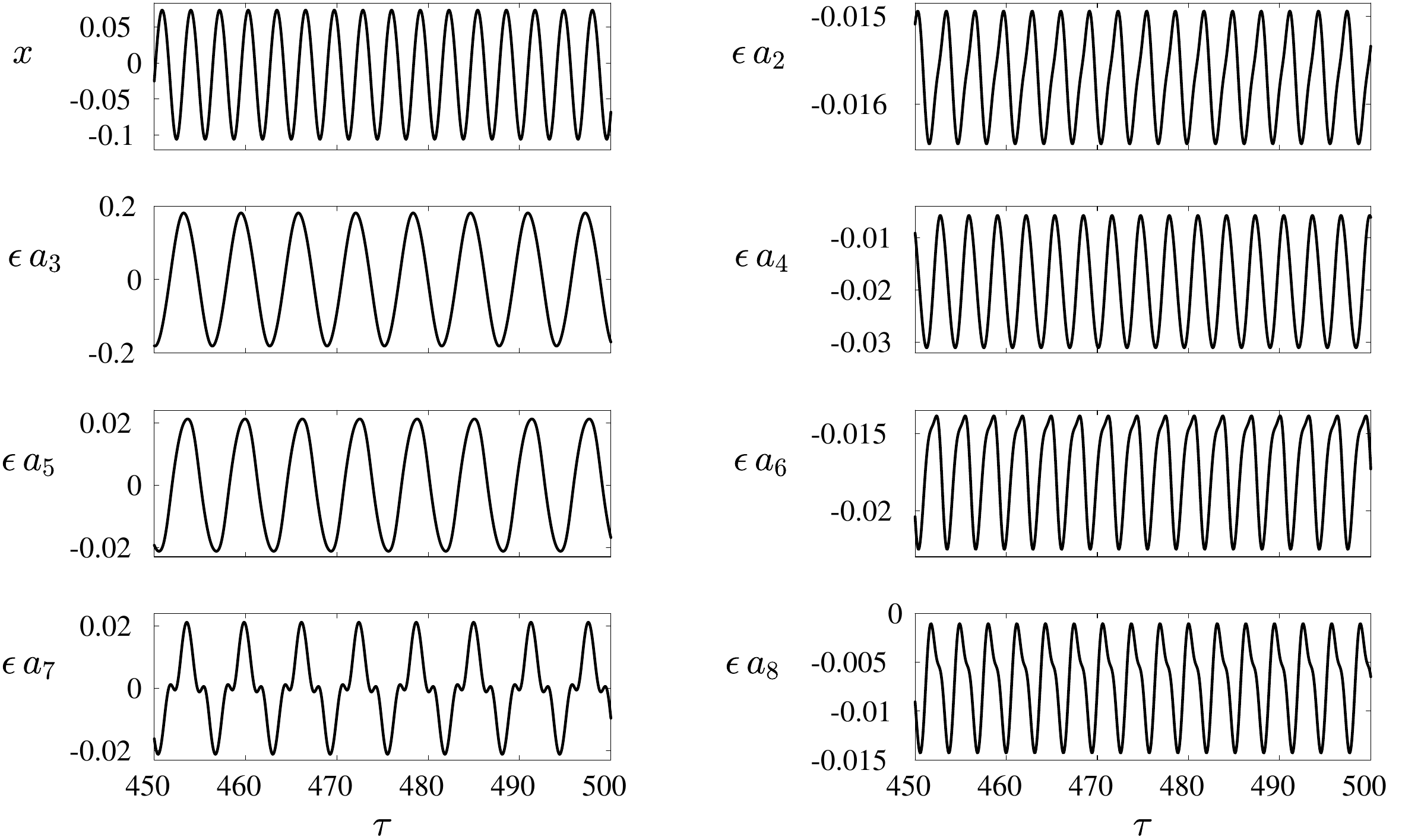}
    \caption{Steady state oscillations of the spherical and the seven first shape modes for the EB with $R_0=2\,\mu$m driven above the first parametric resonance of the $n=3$ mode with $p_a=0.92\,$MPa, $f=8.52\,$MHz, and $\Omega_{0,n}\sim1$ (corresponding to the point $\bullet$ in figure~\ref{f:n2stab}) with the set of interface parameters in table~\ref{tb:1}.}
    \label{f:n3}
\end{figure}

In this section, the variation of the spherical and shape mode amplitudes for an EB are discussed in detail. The coupled governing equations are solved for the interface parameters tabulated in table~\ref{tb:1}. The two quantities $x$ and $\epsilon s_n$, respectively, given by
\begin{align}
    x=\f{R(t)}{R_0}-1, \quad \epsilon s_n=\epsilon \f{a_n(t)}{R_0},
\end{align}
are plotted against time $(t)$. Eight modes are retained in the computations and the initial conditions for all the shape modes are set to $s_n(0)=10^{-2}$ and $\dot{s}_n(0)=0$. Prior to these calculations the steady state simulation of the Rayleigh-Plesset radial equation has been run to set the initial conditions for $R(0)$ and $\dot{R}(0)$. From the direct numerical simulations of the coupled governing equations it is observed that the parametrically forced even mode $(n=2)$ only excites the even modes whilst the odd modes $(n=3)$ can excite both even and odd modes, as shown in figures~\ref{f:n2} and \ref{f:n3}, respectively. This aligns with that of the observations highlighted by \citet{Shaw2006} and \citet{guedra2018bubble}. However, it is important to emphasize that the interface parameters play a crucial role in this. 

For the case of $n=2$, specific observations can be made regarding the consequences of setting certain interface parameters to zero. This enables us to interpret the effect of the specific interface parameter in the numerical simulations while keeping all other working parameters unchanged. 
When the interface parameter $\gamma_{3}$, which captures the curvature effects, is set to zero, the even shape modes still get excited. However, the amplitude of shape mode oscillations is slightly higher in this case. When the interface parameter $\gamma_{4}$, which accounts for the initial size-dependent effects, is set to zero, it leads to all the shape modes being zero. The dominant nature of the initial size of the bubble in finite shape mode oscillations can be attributed to the fact that the radius of the EB directly influences its natural frequency of oscillation. Smaller EBs have higher natural frequencies, making them more responsive to external perturbations, resulting in more noticeable shape mode oscillations. When interface parameter $\gamma_{5}$ is set to zero, the even shape mode exhibit a finite amplitude, though with slightly smaller amplitudes. It is also observed that the interface parameters $(\gamma_1,\gamma_2,\gamma_4)$ play a more significant role compared to $(\gamma_3,\gamma_5)$. This aspect can also be understood by looking at their net contributions in the expressions of effective interface tension parameters in \eqref{e:sig}. The net effect of $(\gamma_1,\gamma_2,\gamma_4)$ at both the inner and outer interfaces gets added up since $\gamma_i=(\gamma_{1i}+\gamma_{2i}),\,i=\{1,\,2,\,4\}$, resulting in the increase of their net contributions. Whereas in the case of $(\gamma_3,\gamma_5)$, the outer interface possesses a negative sign due to the orientation of the normal vectors, such that $\gamma_{j}=(\gamma_{1j}-\gamma_{2j}),\,j=\{3,\,5\}$, resulting in the decrease of their net contributions.

\begin{figure}
    \centering
    \includegraphics[width=\linewidth]{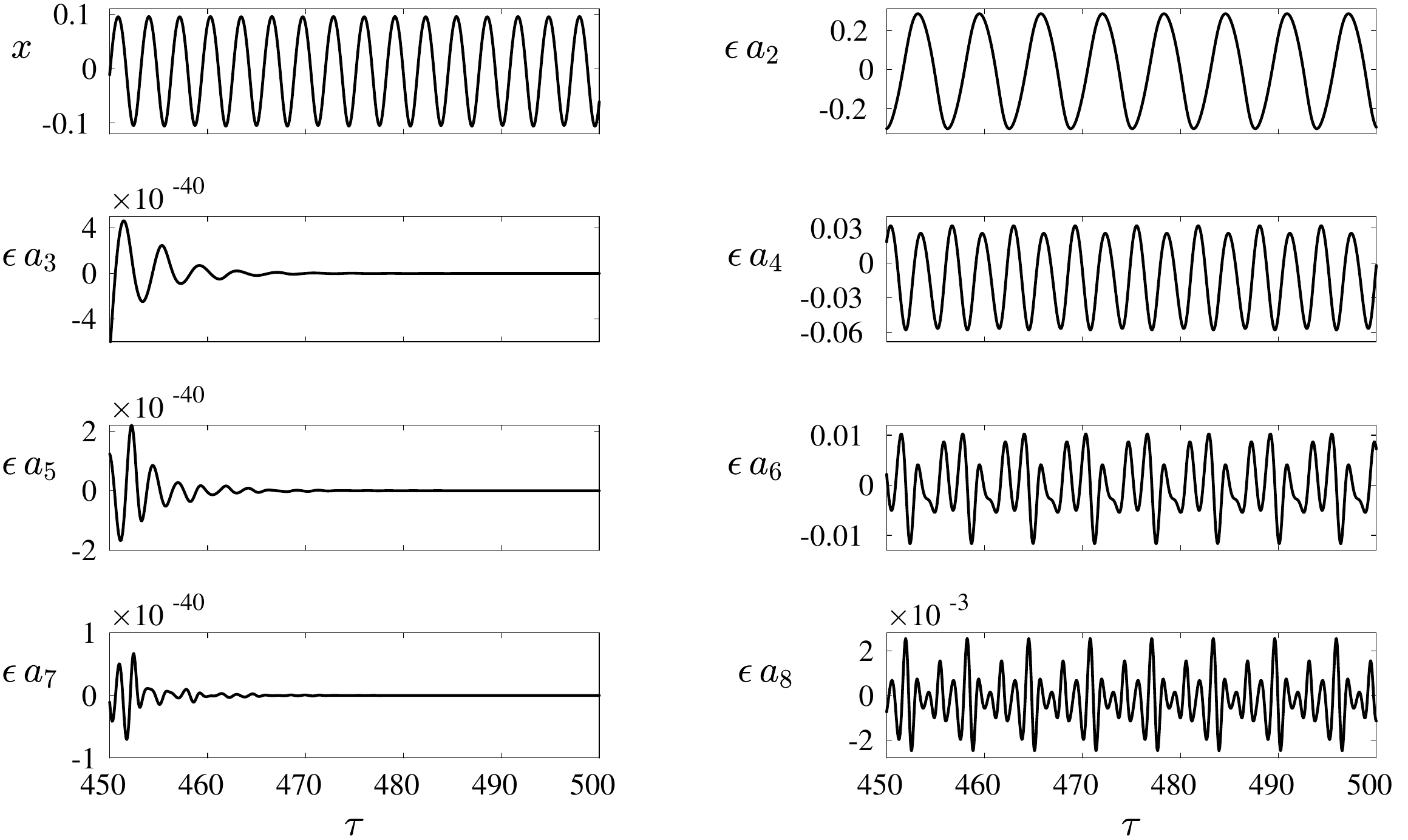}
    \caption{Steady state oscillations of the spherical and the seven first shape modes for the EB with $R_0=5\,\mu$m driven above the first parametric resonance of the $n=2$ mode with $p_a=0.07\,$MPa, $f=1.17\,$MHz, and $\Omega_{0,n}\sim1$ (corresponding to the point $\bullet$ in figure~\ref{f:r5n2stab}) with the set of interface parameters in table~\ref{tb:1}.}
    \label{f:r5n2}
\end{figure}

Moreover, we extend this analysis for a relatively larger EB radii to enable us to discuss the influence and significance of interface parameters. The steady state oscillation of spherical and first seven shape modes for an EB with radius $R_0=5\,\mu$m is shown in figure~\ref{f:r5n2}. 
Also as the bubble's radius increases, the importance of interface parameters diminishes and the analysis is predominantly governed by the initial size of the bubble and other physical parameters (like viscosity and inside gas pressure), making it easier to simulate the equations without encountering numerical difficulties. However, for smaller radius bubbles, where a small length scale is involved, the interface parameters exhibit a dominant behavior in conjunction with other physical parameters. Therefore, it is essential to consider their effects, particularly while analysing small radii bubbles. Among other physical parameters, the viscosity and elastic material constant of the shell have consistently played a crucial role in the behavior of EBs, influencing both their radial and nonspherical oscillations. In the present model, their importance remains prominent.

\section{Perturbation analysis using Krylov-Bogoliubov averaging method}
\label{sec:per}

In this section, the Krylov-Bogoliubov asymptotic perturbation technique is used to analyze the system further. By canceling out the secular terms in the coupled governing equations, a set of first-order differential equations is derived. These equations capture the essential dynamics of the system and provide a foundation for further analysis. In the subsequent section, these equations are utilized to perform a steady-state analysis and investigate the conditional stability of the present model. 

Let us represent the spherical mode $R(t)$ and the shape mode $a_n(t)$ in terms of non-dimensional coordinates $x(t)$ and $s_n(t)$, respectively, such that
\begin{equation}
R(t)=R_0[1+x(t)], \quad a_n(t)=R_0s_n(t).
\end{equation}
Additionally, the new time scale $\tau$ is defined as
\begin{equation}
    \tau=\f{\omega t}{2}.
\end{equation}

In this analysis, the non-dimensional coordinate $x$, pressure amplitude $p_a$, viscous damping measuring parameter of surrounding liquid $(\delta/R_0)^2$, and shell $\nu^{\rm S}$ are all considered as small quantities of $\epsilon$ order. By performing a power series expansion of $x$ and considering terms up to the second orders in $\epsilon$, the equations \eqref{e:smode} and \eqref{e:nsmode} can expressed as
\begin{align}
    \ddot x &+\Omega_0^2x=A \cos{2\tau}+\epsilon \mathcal{F}_1 \label{e:smodenew}\\ 
    \ddot s_n&+\Omega_{0,n}^2s_n=\epsilon \mathcal{F}_2.\label{e:nsmodenew}
\end{align}
where $\Omega_0^2=\omega_0^2+4B$, $\Omega_{0,n}^2=\omega_{0,n}^2+4B(n+1)$, $\mathcal{F}_1$ and $\mathcal{F}_2$ are, respectively, given by 
\allowdisplaybreaks{
\begin{align*}
    \mathcal{F}_1&= \bigg[\left(Q+22B\right)x^2-\f{3}{2}\dot x^2-xA\cos{2\tau}-F_{\nu,0}^{(0)}\dot x\bigg]-\f{1}{\mathcal{H}_n}\sum_{n=2}^\infty{\left(n+\f{3}{2}\right)\dot s_n^2} \nn\\
    &+\f{1}{\mathcal{H}_n}\sum_{n=2}^\infty{\left[(n+3)\Omega_{0,n}^2-(n+1)U-(n+1)(n(n+1)-10)B\right]s_n^2}, \nn\\
    \mathcal{F}_2&=\Bigg[(n-1)A\cos{2\tau}+\bigg\{3\Omega_{0,n}^2-(n-1)\Omega_0^2+2(n+1)\big\{14-n(n+1)\big\}B\nn\\
    &-(n+1)I_{\rm p}\bigg\}xs_n-3\dot x \dot s_n-F_{\nu,0}^{(n)}\dot s_n\Bigg] +\mathcal{H}_n\sum_{i=2}^\infty \sum_{j=2}^\infty\Bigg[\Bigg\{\Omega_{0,j}^2\f{M_{c_{nij}}}{4}+W_0 I_{a_{nij}}\nn\\
    &+B\left(5I_{a_{nij}}+I_{c_{ijn}}+\f{I_{c_{nij}}}{2}\right)+N_{nij}\Bigg\}s_i s_j-\f{M_{d_{nij}}}{4}\dot s_i \dot s_j\Bigg].
\end{align*}
}
Here $\dot{(\cdot)}$ represents the time derivatives with respect to $\tau$. The other parameters in \eqref{e:smodenew} and \eqref{e:nsmodenew} are given by
{\allowdisplaybreaks
\begin{subequations}
    \label{e:ndexp}
\begin{align}
    A&=\left(\f{2}{\omega}\right)^2 \f{p_a}{\rho^{\rm L} R_0^2},\quad
    B=\left(\f{2}{\omega}\right)^2 \f{6C_1q}{\rho^{\rm L} R_0^3},\quad
    W_0=\left(\f{2}{\omega}\right)^2 \f{\sigma_{\rm eq}}{\rho^{\rm L} R_0^3}, \label{e:Apar}\\
    W&=\left(\f{2}{\omega}\right)^2 \f{1}{\rho^{\rm L} R_0^3}\Bigg[(n-1)(n+2)\bigg\{\gamma_0+2\gamma_1+2\gamma_2  +5\gamma_4\f{1}{R_0^2}+2\gamma_3\f{1}{R_0}+8\gamma_5\f{1}{R_0}\bigg\} \nn\\
    &\qquad \qquad  +4\gamma_2+8\gamma_4\f{1}{R_0^2}+16\gamma_5\f{1}{R_0}\Bigg],\\
    Q&=\left(\f{2}{\omega}\right)^2\f{1}{\rho^{\rm L} R_0^2}\Bigg[\f{9}{2}k(1+k)p_{g_0}-\f{2}{R_0}\bigg\{2\gamma_0+4\gamma_1-2\gamma_2 +5\gamma_3\f{1}{R_0}+2\gamma_4\f{1}{R_0^2}\bigg\}\Bigg],\\
    U&=\left(\f{2}{\omega}\right)^2 \f{1}{\rho^{\rm L} R_0^2}\Bigg[3kp_{g_0}-\f{2}{R_0}\bigg\{\gamma_0+2\gamma_1 -(n(n+1)+4)\gamma_2 +\left(\f{n(n+1)+2}{2}\right)\gamma_3\f{1}{R_0}\nn\\
    &\qquad\qquad+\gamma_4\f{1}{R_0^2}-2n(n+1)\gamma_5\f{1}{R_0}\bigg\} \Bigg],\\
    I_{\rm p}&=\left(\f{2}{\omega}\right)^2 \f{1}{\rho^{\rm L} R_0^3}\Bigg[(n-1)(n+2)\bigg\{4\gamma_2-2\gamma_3\f{1}{R_0} +8\gamma_5\f{1}{R_0}\bigg\}+8\gamma_2+16\gamma_5\f{1}{R_0}\Bigg],\\
    \omega_0^2&=\left(\f{2}{\omega}\right)^2 \f{1}{\rho^{\rm L} R_0^2}\Bigg[3kp_{g_0}-\f{2}{R_0}\bigg\{\gamma_0+2\gamma_1-2\gamma_2+2\gamma_3\f{1}{R_0}+\gamma_4\f{1}{R_0^2}\bigg\}\Bigg],\\
    \omega_{0,n}^2&=(n+1)W,\\
    N_{nij}&=\left(\f{2}{\omega}\right)^2 \f{1}{2\rho^{\rm L} R_0^3}\Bigg[N^2_{nij}\gamma_2+N^3_{nij}\gamma_3\f{1}{R_0}+N^4_{nij}\gamma_4\f{1}{R_0^2}+N^5_{nij}\gamma_5\f{1}{R_0}\Bigg].
\end{align}
\end{subequations}
}
The functions representing the viscous dissipation in the spherical $F_{\nu,0}^{(0)}$ and nonspherical $F_{\nu,0}^{(n)}$ mode equations, at the lowest order, are given by
\begin{subequations}
\begin{align}
    F_{\nu,0}^{(0)}&=\left(\f{2}{\omega}\right)\f{4\nu^{\rm S}}{R_0^2},\\
    F_{\nu,0}^{(n)}&=\left(\f{2}{\omega}\right)\f{2}{R_0^2}\left[(n+2)\left\{(2n+1)\nu^{\rm S}-2n(n+2)\f{\delta}{R_0}\nu^{\rm L}\right\}\right].
\end{align}
\end{subequations}
Consistent with the discussions in \citet{guedra2018bubble}, Krylov-Bogoliubov averaging method \citep{nayfeh2008perturbation} is followed to find the solutions to \eqref{e:smodenew} and \eqref{e:nsmodenew}. In order to obtain the solutions of \eqref{e:smodenew} and \eqref{e:nsmodenew}, the following form of asymptotic expansions for $x(\tau)$ and $s_n(\tau)$, respectively, are assumed
\begin{align}
    x(\tau)&=x_0+\epsilon x_1+\epsilon^2 x_2 + \cdot \cdot \cdot \;\;, \label{e:xtau}\\
    s_n(\tau)&=s_{n,0}+\epsilon s_{n,1} + \epsilon^2 s_{n,2} + \cdot \cdot \cdot \;\;. \label{e:sntau}
\end{align}

For a shape mode $n$ close to a parametric resonance, such that $\Omega_{0,n}\sim a$ ($a$ is an integer), the solution to $\mathcal{O}(\epsilon^0)$ equation of \eqref{e:nsmodenew} can be assumed to be of the form
\begin{align}
    s_{n,0}=S_n \cos{\theta_n},\quad \theta_n=a\tau+\phi_n. \label{e:sn0}
\end{align}
In this analysis, the first parametric resonance of the shape mode $n$ is considered exclusively, hence $a=1$. In the case of spherical oscillations of the bubble, it is assumed that the oscillations are significantly far from the harmonic resonances, ensuring that $\Omega_0 \neq 2a$. Considering that the steady state of the spherical oscillations has already been achieved at the initial time, the solution to \eqref{e:smodenew} in the non-resonant region can be assumed to be of the form
\begin{align}
    x_0=X\cos{\theta_x},\quad \theta_x=2\tau+\phi_x. 
\label{e:x0}
\end{align}
To account for the dynamics of other modes where $m\neq n$, it is assumed that these secondary modes are sufficiently far from their parametric resonances $\Omega_{0,m} \neq a)$ and that these $m$ modes oscillate at the frequency of the excitation forcing at the lowest order, such that
\begin{align}
    s_{m,0}=S_m \cos{\theta_m}, \quad \theta_m=2\tau+\phi_m \;\; (m \neq n). 
\label{e:sm0}
\end{align}

In order to apply Krylov-Bogoliubov method, the amplitudes $(X, S_n, S_m)$ and phases $(\phi_x, \phi_n, \phi_m)$ are assumed to exhibit slow variations over time $\tau$. Also, assuming that 
$$\dot{x}_0=-X\sin{\theta_x},\quad\dot{s}_{n,0}=-S_n \sin{\theta_n},\quad \dot{s}_{m,0}={S}_m\cos{\theta_m}$$ leads to the following three set of equations 
\begin{align}
    \dot{X}\cos\theta_x-X\sin \theta_x \dot\phi_x=0,\nn\\
    \dot{S}_n\cos\theta_n-S_n\sin \theta_n \dot\phi_n=0,\label{e:first}\\
    \dot{S}_m\cos\theta_m-S_m\sin \theta_m \dot\phi_m=0.\nn
\end{align}

Substituting \eqref{e:sn0}, \eqref{e:x0} and \eqref{e:sm0} back into \eqref{e:smodenew} and \eqref{e:nsmodenew} leads to the following equations
{\allowdisplaybreaks
\begin{subequations}
\label{e:second}
\begin{align}
    &-2\dot{X}\sin\theta_x-2X\cos\theta_x(2+\dot\phi_x)+\Omega_0^2X\cos\theta_x=A\cos2\tau+\epsilon \mathcal{F}_1(Z),\\
    &-\dot S_n\sin\theta_n-S_n\cos\theta_n(1+\dot\phi_n)+\Omega_{0,n}^2 S_n\cos\theta_n=\epsilon \mathcal{F}_2(Z),\\
    &-2\dot{S}_m\sin\theta_m-2S_m\cos\theta_m(2+\dot\phi_m)+\Omega_{0,m}^2S_m\cos\theta_m=\epsilon \mathcal{F}_2(Z).
\end{align}
\end{subequations}
} 
where $Z=\{X\cos\theta_x,S_n\cos\theta_n,S_m\cos\theta_m, -X\sin\theta_x,-S_n\sin\theta_n,-S_m\sin\theta_m\}$. Solving the system of equations \eqref{e:first} and \eqref{e:second} leads to six set of first order ordinary differential equations in terms of $(X, \phi_x, S_n, \phi_n, S_m, \phi_m)$ as
{\allowdisplaybreaks
\begin{subequations}
\label{e:bavg}
\begin{align}
    -2\dot X&=\sin\theta_x \Big[(4-\Omega_0^2)X\cos\theta_x+A\cos2\tau+\epsilon \mathcal{F}_1(Z)\Big],\\
    -2X\dot \phi_x&=\cos\theta_x \Big[(4-\Omega_0^2)X\cos\theta_x+A\cos2\tau+\epsilon \mathcal{F}_1(Z)\Big],\\
    -\dot S_n&=\sin\theta_n\Big[\big(1-\Omega_{0,n}^2\big)S_n\cos\theta_n+\epsilon \mathcal{F}_2(Z)\Big],\\
    -S_n\dot \phi_n &=\cos\theta_n\Big[\big(1-\Omega_{0,n}^2\big)S_n\cos\theta_n+\epsilon \mathcal{F}_2(Z)\Big],\\
    -2\dot S_m&=\sin\theta_m\Big[\big(4-\Omega_{0,m}^2\big)S_m\cos\theta_m+\epsilon \mathcal{F}_2(Z)\Big],\\
    -2S_m \dot \phi_m&=\cos\theta_m\Big[\big(4-\Omega_{0,m}^2\big)S_n\cos\theta_m+\epsilon \mathcal{F}_2(Z)\Big].
\end{align}
\end{subequations}
} 
The first approximation of the Krylov-Bogoliubov begins with the Fourier expansion of the right hand side of \eqref{e:bavg} in terms of $\theta_x, \theta_n$ and $\theta_m$ and identifying the leading order term. These terms are also called as averaged equations or slow-time equations given by 
{\allowdisplaybreaks
\begin{subequations}
\label{e:fodeq}
\begin{align}
    -4\dot X&=2F_{\nu,0}^{(0)}+A\sin{\phi_x}+S_n^2\f{\beta_n}{2}\sin{(\phi_x-2\phi_n)},\\
    -4X\dot \phi_x&=(4-\Omega_0^2)X+A\cos{\phi_x}+S_n^2\f{\beta_n}{2}\cos{(\phi_x-2\phi_n)},\\
    -2\dot S_n&=F_{\nu,0}^{(n)}S_n+\f{S_n}{2}\Bigg[(n-1)A\sin{2\phi_n}+\zeta_n X\sin{(2\phi_n-\phi_x)}\nn\\
    &\quad+\sum_{m\neq n} W_{nm} S_m\sin{(2\phi_n-\phi_m)}\Bigg]\\
    -2S_n\dot \phi_n &=(1-\Omega_{0,n}^2)S_n+\f{S_n}{2}\Bigg[(n-1)A\cos{2\phi_n}+\zeta_n X \cos{(2\phi_n-\phi_x)} \nn\\
    &\quad +\sum_{m\neq n} W_{nm} S_m \cos{(2\phi_n-\phi_m)}\Bigg],\\
    -4\dot S_m&=2F_{\nu,0}^{(m)}S_m+ O_{mn} \f{S_n^2}{2}\sin{(\phi_m-2\phi_n)},\\
    -4S_m \dot \phi_m&=(4-\Omega_{0,m}^2)S_m+ O_{mn} \f{S_n^2}{2}\cos{(\phi_m-2\phi_n)},
\end{align}
\end{subequations}
} 
\noindent where the coefficients $\beta_n,\zeta_n,W_{nm},O_{mn}$ are listed in Appendix~\ref{app:B}.
In the subsequent section, the first-order differential equations in \eqref{e:fodeq} are used to provide the steady-state solutions and derive analytical expressions for the amplitude of shape oscillations.

\section{Steady-state solutions and stability analysis}
\label{sec:stab}

The set of differential equations in \eqref{e:fodeq} is written in terms of a complex form by introducing the complex variables $\bar{X}=X\text{e}^{\text{i}\phi_x}$, $\bar{S}_n=S_n\text{e}^{\text{i}\phi_n}$ and $\bar{S}_m=X\text{e}^{\text{i}\phi_m}$ to get \citep{guedra2018bubble}
{\allowdisplaybreaks
\begin{subequations}
\begin{align}
    4\text{i}\dot{\bar X}&=\bar\Delta_0 \bar X+A+\f{\beta_n}{2} \bar {S}_n^2, \label{e:compx} \\
    2\text{i}\dot{\bar{S}}_n&=\bar\Delta_n \bar{S}_n+\left[\zeta_n \bar X +(n-1)A+\sum_{m\neq n}W_{nm} \bar{S}_m\right]\f{\bar{S}_n^*}{2}, \label{e:compsn} \\
    4\text{i}\dot{\bar{S}}_m&=\bar\Delta_m \bar{S}_m+O_{mn}\f{\bar{S}_n^2}{2}, \label{e:compsm}
\end{align}
\end{subequations}
}
where $*$ denotes the conjugate, and the complex quantities $\bar\Delta_{(\cdot)}$ are given by
{\allowdisplaybreaks
\begin{subequations}
\begin{align}
    \bar\Delta_0&=4-\Omega_0^2-2\text{i}F_{v,0}^{(0)}, \label{e:del0} \\
    \bar\Delta_n&=1-\Omega_{0,n}^2-\text{i}F_{v,0}^{(n)}, \label{e:deln} \\
    \bar\Delta_m&=4-\Omega_{0,m}^2-2\text{i}F_{v,0}^{(m)}. \label{e:delm}
\end{align}
\end{subequations}
}
In the steady-state regime, the first temporal derivatives in equations \eqref{e:compx} and \eqref{e:compsm} can be cancelled and the complex amplitudes of the spherical and secondary shape modes can directly be determined as
\begin{align}
    \bar X &=-\bar{\Delta}_0^{-1}\left(A+\f{\beta_n}{2}\bar{S}_n^2\right), \label{e:xbar}\\
    \bar{S}_m &=-\bar{\Delta}_m^{-1}O_{mn}\f{\bar{S}_n^2}{2}.\label{e:smbar}
\end{align}
Further, the above relations in \eqref{e:xbar} and \eqref{e:smbar} are substituted in \eqref{e:compsn} to get
\begin{align}
    \left[2\bar{\Delta}_n-\f{S_n^2}{2}\left(\f{\zeta_n\beta_n}{\bar{\Delta}_0}+\sum_{m \neq n}\f{W_{nm} O_{mn}}{\bar{\Delta}_m}\right)\right]\bar{S}_n=-\left[(n-1)-\f{\zeta_n}{\bar{\Delta}_0}\right]A \bar{S}_n^*.
\end{align}
The trivial solution for the above equation is $S_n=0$, such that
\begin{align}
    \bigg\rvert \bar{y}_1-\f{S_n^2}{2}\bar{y}_2\bigg\rvert =\rvert \bar{y}_3 \rvert A,\label{e:ycon1} \\
    \text{e}^{2\text{i}\phi_{n}}=-\f{\bar{y}_3A}{\bar{y}_1-\cfrac{S_n^2}{2}\bar{y}_2}.\label{e:ycon2}
\end{align}
Here the complex quantities $\bar{y}_{(\cdot)}$ are defined as
\begin{align}
    \bar{y}_1 &= 2\bar{\Delta}_n, \label{e:y1bar} \\
    \bar{y}_2 &= \bar{\Delta}_0^{-1} \zeta_n \beta_n +\sum_{m \neq n} \bar{\Delta}_m^{-1} W_{nm} O_{mn},\label{e:y2bar}\\
    \bar{y}_3 &= (n-1)-\bar{\Delta}_0^{-1} \zeta_n.\label{e:y3bar}
\end{align}
It is worth mentioning that the solutions obtained in this analysis include the onset threshold for the parametric excitation of the shape mode $n$ in the vicinity of the first resonance. Following a similar approach as discussed in \citet{guedra2018bubble}, an analytical expression for the absolute stability threshold (in terms of driving amplitude pressure) is derived by neglecting the quadratic terms in \eqref{e:xbar} and \eqref{e:ycon1}, which can be written as
\begin{align}
    A_{\text{th}}=\Bigg\rvert \f{\bar{y}_1}{\bar{y}_3} \Bigg\rvert = \cfrac{2\rvert \bar{\Delta}_n\rvert \rvert \bar{\Delta}_0\rvert}{\rvert\zeta_n -(n-1)\bar{\Delta}_0\rvert}.
\label{e:Ath}
\end{align}
Similarly, an expression can also be deduced in terms of the spherical mode driving amplitude
\begin{align}
    X_{\text{th}}=\cfrac{2\rvert \bar{\Delta}_n\rvert}{\rvert\zeta_n -(n-1)\bar{\Delta}_0\rvert}.
\label{e:Xth}
\end{align}

\begin{figure}
    \centering
    \includegraphics[width=0.99\linewidth]{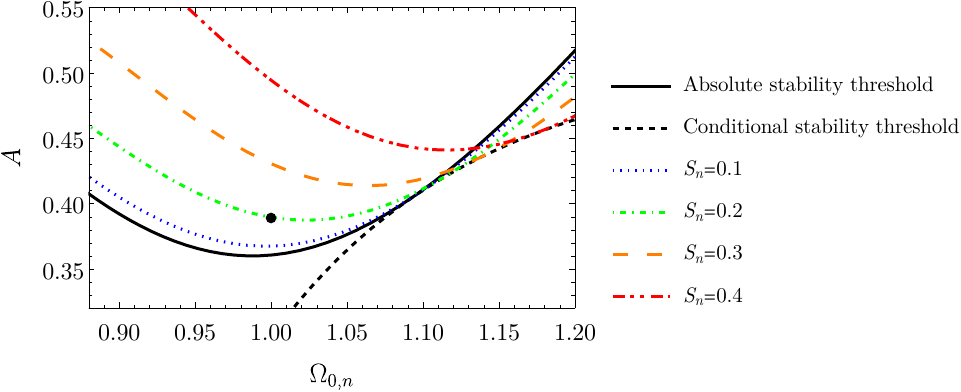}
    \caption{The plot shows the driving amplitude $A$ as a function of the frequency $\Omega_{0,n}$, and $n=2$ for the EB with $R_0=2\,\mu$m. Curves corresponds to a specific stability threshold and fixed values of $S_n$ in the steady state with the set of interface parameters in table~\ref{tb:1}.}
    \label{f:n2stab}
\end{figure}

Upon comparing the relations given in \eqref{e:Ath} and \eqref{e:Xth} with those in \citet{guedra2018bubble}, it is evident that the structure of these relations remains unchanged. However, the definitions of the involved quantities differ due to the inclusion of interface parameters and the shell elasticity constant in the present model. 
The steady-state solutions of \eqref{e:ycon1} can be written as
\begin{align}
    S_n=\f{\sqrt{2h_0}}{y_2}\left[1\pm\sqrt{1+\cfrac{y_2^2(y_3A-y_1)(y_3A+y_1)}{h_0^2}} \right]^{1/2}.
\end{align}
Here $y_{i}=\rvert \bar{y}_i \rvert$, and
\begin{align}
    h_0=\Re(\bar{y}_1)\Re(\bar{y}_2)+\Im(\bar{y}_1)\Im(\bar{y}_2).
\end{align}
The conditional stability threshold $(A'_{\text{th}})$ is the condition for $A$ that ensures the solutions are real $(A>0)$, then
\begin{align}
    A>A^{'}_{\text{th}}=\f{1}{y_3}\sqrt{\left(y_1-\f{h_0}{y_2}\right)\left(y_1+\f{h_0}{y_2}\right)}.
\label{e:Athdash}
\end{align}

\begin{figure}
    \centering
    \includegraphics[width=0.99\linewidth]{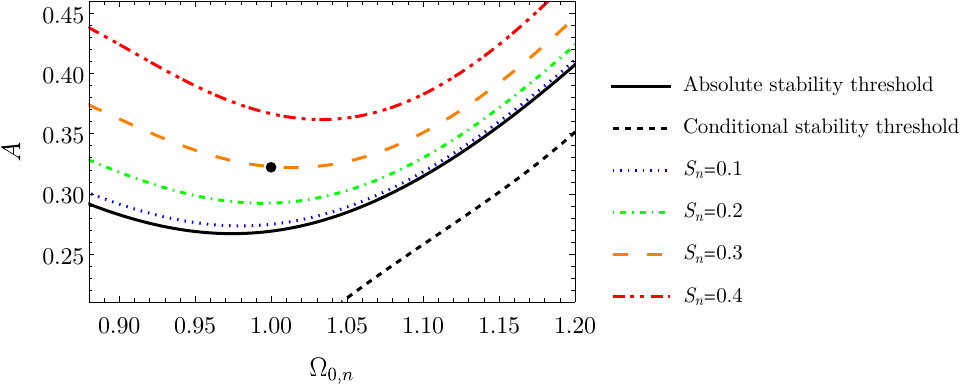}
    \caption{The plot shows the driving amplitude $A$ as a function of the frequency $\Omega_{0,n}$, and $n=3$ for the EB with $R_0=2\,\mu$m. Curves corresponds to a specific stability threshold and fixed values of $S_n$ in the steady state with the set of interface parameters in table~\ref{tb:1}.}
    \label{f:n3stab}
\end{figure}

The driving amplitude $(A)$ corresponding to isovalues of shape mode amplitude $(S_n)$ can directly be calculated using the relation in \eqref{e:ycon1}. 
The absolute stability threshold $(A_{\text{th}})$ and the conditional stability threshold $(A'_{\text{th}})$ can be calculated from \eqref{e:Ath} and \eqref{e:Athdash}, respectively. The driving amplitudes for various values of $S_n$ (isolines), along with the absolute and conditional stability thresholds, are plotted against $\Omega_{0,n}$ for $n=2$ and $n=3$ in figures~\ref{f:n2stab} and \ref{f:n3stab}, respectively.

In figure~\ref{f:n2stab}, the point marked with $(\bullet)$ on the isoline $S_n=0.2$ represents the value of $A$ corresponding to $\Omega_{0,n}\sim 1$. By using the definitions of $A$ and $\omega_{0,n}^2$ as given in \eqref{e:ndexp}, the excitation pressure and frequency values are determined for this specific point. Subsequently, these calculated pressure and frequency values are considered in the direct numerical simulations, which lead to the determination of the finite shape mode amplitudes illustrated in figure~\ref{f:n2}. It is also apparent that the amplitude of oscillation of shape mode $\epsilon a_2=0.2$ corresponds to the isoline $S_n=0.2$.

A similar analysis is conducted for $n=3$, at the point denoted by $(\bullet)$ on the isoline $S_n=0.3$ in figure~\ref{f:n3stab}, and the calculated values of pressures and frequency is used in the direct numerical simulations depicted in figure~\ref{f:n3}. It is observed that a slight change in the excitation pressure would still yield in finite amplitude oscillations in the time-series analysis, while a significant change would cause the finite non-zero amplitudes to drop to zero. To study the temporal variation, it is crucial to ensure that the simulations are conducted within and around the conditional stable zone close to the corresponding $\Omega_{0,n}\sim 1$. Hence, for a specific value of $n$, conditional stability plots can be generated, and based on those, appropriate working values of excitation pressure and frequency are calculated at $\Omega_{0,n}\sim 1$ to conduct further temporal evolution analysis. 

\begin{figure}
    \centering
    \includegraphics[width=0.99\linewidth]{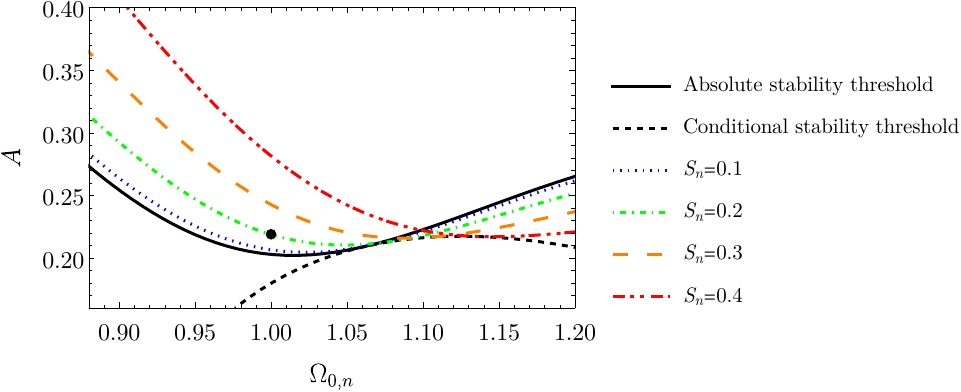}
    \caption{The plot shows the driving amplitude $A$ as a function of the frequency $\Omega_{0,n}$, and $n=2$ for the EB with $R_0=5\,\mu$m. Curves corresponds to a specific stability threshold and fixed values of $S_n$ in the steady state with the set of interface parameters in table~\ref{tb:1}.}
    \label{f:r5n2stab}
\end{figure}

As an extended analysis, we consider slightly larger radii bubbles with an initial radius of $R_{0}=5\,\mu$m and a shell thickness of $q=20\,$nm. The stability curves for $n=2$ using the same interface material parameters as before are illustrated in figure~\ref{f:r5n2stab}. Subsequently, direct numerical simulations are conducted for this case, using the calculated values of excitation pressure and frequency, as depicted in figure~\ref{f:r5n2}. Although the analysis remains unchanged, the larger bubble radius leads to a decrease in the effects of interface parameters.

It is essential to emphasize that the direct numerical simulations of the governing equations are influenced by the choice of numerical values of interface parameters and the viscosity of the shell. In certain cases, these simulations may exhibit a blow-up of the solution, indicating the presence of a finite time singularity in the problem. One such possibility is when there are no interface parameters. The blow-up of the solution can also be interpreted as an unstable configuration of the EB, because it represents a sudden and uncontrollable escalation of the bubble's oscillations, which can lead to unpredictable behavior and potential collapse of the bubble. The introduction of interface effects through the surface tension parameter plays an important role in averting this blow-up. However, it's crucial to carefully balance the benefits of these interface parameters against the potential complexity introduced by them in the system. Hence, the behavior of the bubble is strongly affected by these interface parameters, a finding that has also been reported in other interface energy models studying the radial dynamics of EBs.\citep{dash2022describing,dash_tamadapu_2022}
Properly chosen interface parameters effectively address and prevent the occurrence of finite time singularities in the problem, as demonstrated in the current study. The interface parameters modify the EB's behavior, contributing to stability and preventing the undesirable blow-up phenomenon. Additionally, it is worth noting that higher shell viscosity values can lead to the emergence of such finite time singularities and special numerical techniques are required to handle these singularities. 

\section{Conclusion}
\label{sec:con}

This work is focused on developing a mathematical model to investigate the nonspherical oscillations of smaller radii EB suspended in fluid. This model takes into account the essential interfacial mechanics at the gas-encapsulation and encapsulation-fluid interfaces, which contributes significantly to the mechanics of smaller radii bubbles. \citep{dash_tamadapu_2022,dash2022describing} The shell material is treated as a thin membrane with both elastic and viscous effects. The coupled dynamical equations governing the spherical and shape mode oscillations of the EB are derived using Lagrangian energy formulation. These governing equations are then analyzed using direct numerical simulations. 

The study reveals that the parametrically forced even mode $(n=2)$ excites only the even modes, while the odd modes $(n=3)$ excites both even and odd modes. This has also been observed in the larger radii bubbles. \citep{guedra2018bubble} But in the case of smaller radii EBs, the interface parameters play a crucial role in the analysis, unlike in larger bubbles. 
For instance, \citet{guedra2018bubble} reported the finite amplitude oscillations of larger radii bubbles with conventional surface tension parameter without considering the effect of interface energy. But in the present analysis with smaller radii bubbles, we observe that the interface parameters are essential in identifying the stable finite amplitude shape mode oscillations. 
In the absence of interface parameters, the EB demonstrates an unstable behavior, possibly causing solutions to blow up. However, the EB shows finite amplitude oscillations when interface parameters are introduced. The interface energy models describing the radial dynamics of EBs have proven the significant influence of interface parameters.\citep{dash_tamadapu_2022,dash2022describing} The present analysis further highlights this notion, illustrating that interface parameters play an even more substantial and influential role when it comes to nonspherical oscillations. This is substantiated by the fact that finite amplitude shape mode oscillations are observed in smaller radii EBs only when all interface parameters are considered; otherwise, they are not present.
Alongside the interface parameters, the elasticity and viscosity of the shell material are also essential factors in the numerical simulations. Additionally, perturbation analysis has been used to derive the equations and analyze the system in the vicinity of the first parametric resonance of shape mode $n$. The perturbation method based on the Krylov-Bogoliubov method of averaging and steady-state analysis proves invaluable in calculating the required excitation pressure and frequency for the existence of shape mode oscillations. It helps reveal insights into how the resonance and the introduced perturbations affect the EB's stability and amplitude of oscillations. This information is particularly valuable in determining optimal excitation pressures and frequencies for small radii bubbles, which hold significant promise for various medical applications.

Future work includes several directions. The present study can be extended to investigate the nonspherical oscillations of EBs suspended in any biological fluids, which can be assumed as linear or nonlinear viscoelastic fluids. \citep{allen2000dynamics,doi:10.1121/1.429344} In the context of biomedical applications, when EBs approach blood vessels, their dynamic behavior alters. \citep{doinikov2009modeling,suslov2012nonlinear} Another interesting extension is exploring the nonspherical oscillations of EBs near blood vessel walls using the interface energy model. Interesting recent developments in bubble dynamics involve strategies like coating bubbles with magnetic nanoparticles to enhance their targeting efficiency in drug delivery.\citep{stride2009enhancement,jamburidze2019nanoparticle,malvar2018nonlinear} Moreover, by suspending EBs in magnetic fluids, researchers gain insights into how they respond under the combined influence of acoustic and magnetic fields. The interface energy model can also be used to study the behavior of these magnetic microbubbles.

\section*{Acknowledgements}
We are grateful to P. G. Senapathy Center for Computing Resource, IIT Madras, for providing the computational resources. This work was supported by the Science and Engineering Research Board (SERB), Department of Science and Technology, Government of India, under the project number CRG/2022/005775.

\appendix

\section{Basis invariants for the interface}
\label{app:A0}
The interface energy density can be expressed in terms of six basis invariants of the right Cauchy-Green interface deformation tensor $\msb{C}$, the relative curvature tensor $\bs{\kappa}$, and the permutation tensor-density $\bs{\mu}$ on the undeformed interface as
\begin{align}
    \gamma=\gamma(I_1, I_2, I_3, I_4, I_5, I_6),
\end{align}
and the invariants are given by 
\begin{align}
\begin{split}
    I_1&={\rm tr}\,\msb{C}=G^{\alpha\beta}C_{\alpha\beta},\\
    I_2&={\rm det}\,\msb{C}=J^2=g/G,\\
    I_3&={\rm tr}{\,\bs \kappa}=G^{\alpha\beta}\kappa_{\alpha\beta},\\
    I_4&={\rm det}{\,\bs \kappa}=\f{1}{2}\mu^{\alpha\beta}\mu^{\gamma\delta}\kappa_{\alpha\gamma}\kappa_{\beta\delta},\\
    I_5&={\rm tr}(\msb{C}\bs{\kappa})=C_{\alpha\beta}\kappa^{\alpha\beta}=C^{\alpha\beta}\kappa_{\alpha\beta},\\
    I_6&={\rm tr}(\msb{C}\bs{\kappa}\bs{\mu})=G_{\alpha\beta}C_{\gamma\delta}\kappa^{\alpha\gamma}\mu^{\beta\delta}=G_{\alpha\beta}\kappa_{\gamma\delta}C^{\alpha\gamma}\mu^{\beta\delta}.
    \end{split}
\label{e:invariants}
\end{align}

\section{Nonlinear interaction coefficients}
\label{app:A}
The integrals and the nonlinear interaction coefficients are defined through the following where $\mu=\cos{\phi}$:
{\allowdisplaybreaks
\begin{align}
    I_{a_{nij}}&=\int_{-1}^1 P_nP_iP_j \text{d}\mu,\\
    I_{c_{nij}}&=-\int_{-1}^1 (1-\mu^2)P_n \f{\text{d}P_i}{\text{d}\mu}\f{\text{d}P_j}{\text{d}\mu}\text{d}\mu,\\
    I_{g_{nij}}&=-\int_{-1}^1\mu (1-\mu^2)\f{\text{d}P_n}{\text{d}\mu} \f{\text{d}P_i}{\text{d}\mu}\f{\text{d}P_j}{\text{d}\mu}\text{d}\mu,\\
    L_{a_{nij}}&=(n+1)(n+i+1)I_{a_{nij}}+I_{c_{nij}},\\
    L_{b_{nij}}&=\f{1}{2}(n+1)(n^2+4n+4)I_{a_{nij}}+(n+2)I_{c_{inj}},\\
    L_{c_{nij}}&=(n+i+3)(n+i+2)I_{a_{nij}}+I_{c_{nij}}+I_{c_{inj}},\\
    G_{c_{nij}}&=\f{2n^2-n+1}{n+1}I_{a_{nij}}-\f{2}{n+1}\left(L_{b_{nij}}-\f{2L_{a_{nij}}}{i+1}\right),\\
    G_{d_{nij}}&=\f{2L_{c_{nji}}}{(n+1)(j+1)}-\f{L_{b_{jin}}-j(j-1)I_{a_{nij}}}{j+1}-\f{4(n-1)}{n+1}I_{a_{nij}},\\
    G_{e_{nij}}&=\f{\left[n(n-j)+3(j+1)\right]I_{a_{nij}}+I_{c_{nji}}}{(n+1)(j+1)},\\
    M_{a_{nij}}&=G_{d_{ijn}}-G_{c_{nij}}-G_{c_{inj}}-G_{c_{ijn}},\\
    M_{b_{nij}}&=G_{d_{jin}}+G_{d_{ijn}}+2\left(G_{e_{nij}}+G_{e_{jin}}\right)-G_{d_{nij}}-G_{d_{inj}},\\
    M_{c_{nij}}&=G_{e_{nij}}+G_{e_{jin}},\\
    M_{d_{nij}}&=G_{e_{nij}}+G_{e_{jin}}-G_{e_{inj}}\\
    N^2_{nij}&=12 I_{a_{nij}}-2I_{c_{nij}}-4 I_{c_{ijn}},\\
    N^3_{nij}&=\left[\left(\f{n(n+1)}{2}+1\right)I_{c_{nij}}+(i(i+1)+2)I_{c_{ijn}}\right],\\
    N^4_{nij}&=n(n+1) I_{c_{nij}}+2i(i+1)I_{c_{ijn}}-6I_{g_{nij}},\\
    N^5_{nij}&=\{6+i(i+1)+n(n+1)\}I_{a_{nij}}-\left(\f{n(n+1)}{2}+2\right)I_{c_{nij}}\nn\\&\qquad\qquad-(i(i+1)+4)I_{c_{ijn}}+3 I_{g_{nij}}.
\end{align}
}

\section{Coefficients appearing in first-order differential equations}
\label{app:B}
In the first order differential equations \eqref{e:fodeq}, the coefficients are given by:
{\allowdisplaybreaks
\begin{align}
    \beta_n&=\f{(n+3)\Omega_{0,n}^2-(n+1)U+\left(n+\f{3}{2}\right)\dot \theta_n^2-B(n+1)\left[n(n+1)-10\right]}{\mathcal{H}_n}, \\
    \zeta_n&=3\Omega_{0,n}^2-(n-1)\Omega_0^2-3\dot \theta_x \dot \theta_n+2(n+1)\left[14-n(n+1)\right]B-(n+1)I_{\rm p},\\
    {W_{nm}}&=\f{\mathcal{H}_n}{4}\bigg[W_0\left(8I_{a_{nnm}}\right)+\Omega_{0,m}^2 M_{c_{nnm}}+\Omega_{0,n}^2 M_{c_{nmn}}-2\left(M_{d_{nnm}}+M_{d_{nmn}}\right)\nn\\
    & \quad +8B\left(5I_{a_{nnm}}+I_{c_{nmn}}+\f{I_{c_{nnm}}}{2}\right)+{4(N_{nnm}+N_{nmn})}\bigg],\\
    {O_{mn}}&=\f{\mathcal{H}_m}{4}\bigg[W_0\left(4I_{a_{mnn}}\right)+\Omega_{0,n}^2 M_{c_{mnn}}+M_{d_{mnn}}+4B\left(5I_{a_{mnn}}+I_{c_{nnm}}+\f{I_{c_{mnn}}}{2}\right)+{4N_{mnn}}\bigg].
\end{align}
}

\bibliography{references}

\begin{thebibliography}{52}%
\makeatletter
\providecommand \@ifxundefined [1]{%
 \@ifx{#1\undefined}
}%
\providecommand \@ifnum [1]{%
 \ifnum #1\expandafter \@firstoftwo
 \else \expandafter \@secondoftwo
 \fi
}%
\providecommand \@ifx [1]{%
 \ifx #1\expandafter \@firstoftwo
 \else \expandafter \@secondoftwo
 \fi
}%
\providecommand \natexlab [1]{#1}%
\providecommand \enquote  [1]{``#1''}%
\providecommand \bibnamefont  [1]{#1}%
\providecommand \bibfnamefont [1]{#1}%
\providecommand \citenamefont [1]{#1}%
\providecommand \href@noop [0]{\@secondoftwo}%
\providecommand \href [0]{\begingroup \@sanitize@url \@href}%
\providecommand \@href[1]{\@@startlink{#1}\@@href}%
\providecommand \@@href[1]{\endgroup#1\@@endlink}%
\providecommand \@sanitize@url [0]{\catcode `\\12\catcode `\$12\catcode
  `\&12\catcode `\#12\catcode `\^12\catcode `\_12\catcode `\%12\relax}%
\providecommand \@@startlink[1]{}%
\providecommand \@@endlink[0]{}%
\providecommand \url  [0]{\begingroup\@sanitize@url \@url }%
\providecommand \@url [1]{\endgroup\@href {#1}{\urlprefix }}%
\providecommand \urlprefix  [0]{URL }%
\providecommand \Eprint [0]{\href }%
\providecommand \doibase [0]{http://dx.doi.org/}%
\providecommand \selectlanguage [0]{\@gobble}%
\providecommand \bibinfo  [0]{\@secondoftwo}%
\providecommand \bibfield  [0]{\@secondoftwo}%
\providecommand \translation [1]{[#1]}%
\providecommand \BibitemOpen [0]{}%
\providecommand \bibitemStop [0]{}%
\providecommand \bibitemNoStop [0]{.\EOS\space}%
\providecommand \EOS [0]{\spacefactor3000\relax}%
\providecommand \BibitemShut  [1]{\csname bibitem#1\endcsname}%
\let\auto@bib@innerbib\@empty
\bibitem [{\citenamefont {Hoff}(2001)}]{hoff2001acoustic}%
  \BibitemOpen
  \bibfield  {author} {\bibinfo {author} {\bibfnamefont {L.}~\bibnamefont
  {Hoff}},\ }\href@noop {} {\emph {\bibinfo {title} {Acoustic characterization
  of contrast agents for medical ultrasound imaging}}}\ (\bibinfo  {publisher}
  {Springer Science \& Business Media},\ \bibinfo {year} {2001})\BibitemShut
  {NoStop}%
\bibitem [{\citenamefont {Stride}\ and\ \citenamefont
  {Saffari}(2003)}]{Stride2003}%
  \BibitemOpen
  \bibfield  {author} {\bibinfo {author} {\bibfnamefont {E.}~\bibnamefont
  {Stride}}\ and\ \bibinfo {author} {\bibfnamefont {N.}~\bibnamefont
  {Saffari}},\ }\bibfield  {title} {\enquote {\bibinfo {title} {Microbubble
  ultrasound contrast agents: A review},}\ }\href {\doibase
  10.1243/09544110360729072} {\bibfield  {journal} {\bibinfo  {journal} {Proc.
  Inst. Mech. Eng., Part H: J. Eng. Med.}\ }\textbf {\bibinfo {volume} {217}},\
  \bibinfo {pages} {429--447} (\bibinfo {year} {2003})}\BibitemShut {NoStop}%
\bibitem [{\citenamefont {Postema}\ \emph {et~al.}(2004)\citenamefont
  {Postema}, \citenamefont {Van~Wamel}, \citenamefont {Lanc{\'e}e},\ and\
  \citenamefont {De~Jong}}]{postema2004ultrasound}%
  \BibitemOpen
  \bibfield  {author} {\bibinfo {author} {\bibfnamefont {M.}~\bibnamefont
  {Postema}}, \bibinfo {author} {\bibfnamefont {A.}~\bibnamefont {Van~Wamel}},
  \bibinfo {author} {\bibfnamefont {C.~T.}\ \bibnamefont {Lanc{\'e}e}}, \ and\
  \bibinfo {author} {\bibfnamefont {N.}~\bibnamefont {De~Jong}},\ }\bibfield
  {title} {\enquote {\bibinfo {title} {Ultrasound-induced encapsulated
  microbubble phenomena},}\ }\href {\doibase
  10.1016/j.ultrasmedbio.2004.02.010} {\bibfield  {journal} {\bibinfo
  {journal} {Ultrasound Med. Biol.}\ }\textbf {\bibinfo {volume} {30}},\
  \bibinfo {pages} {827--840} (\bibinfo {year} {2004})}\BibitemShut {NoStop}%
\bibitem [{\citenamefont {Lindner}(2004)}]{Lindner2004}%
  \BibitemOpen
  \bibfield  {author} {\bibinfo {author} {\bibfnamefont {J.~R.}\ \bibnamefont
  {Lindner}},\ }\bibfield  {title} {\enquote {\bibinfo {title} {Microbubbles in
  medical imaging: current applications and future directions},}\ }\href
  {\doibase 10.1038/nrd1417} {\bibfield  {journal} {\bibinfo  {journal} {Nat.
  Rev. Drug Discovery}\ }\textbf {\bibinfo {volume} {3}},\ \bibinfo {pages}
  {527--533} (\bibinfo {year} {2004})}\BibitemShut {NoStop}%
\bibitem [{\citenamefont {Klibanov}(2006)}]{Klibanov2006}%
  \BibitemOpen
  \bibfield  {author} {\bibinfo {author} {\bibfnamefont {A.~L.}\ \bibnamefont
  {Klibanov}},\ }\bibfield  {title} {\enquote {\bibinfo {title} {Microbubble
  contrast agents},}\ }\href {\doibase 10.1097/01.rli.0000199292.88189.0f}
  {\bibfield  {journal} {\bibinfo  {journal} {Invest. Radiol.}\ }\textbf
  {\bibinfo {volume} {41}},\ \bibinfo {pages} {354--362} (\bibinfo {year}
  {2006})}\BibitemShut {NoStop}%
\bibitem [{\citenamefont {Errico}\ \emph {et~al.}(2015)\citenamefont {Errico},
  \citenamefont {Pierre}, \citenamefont {Pezet}, \citenamefont {Desailly},
  \citenamefont {Lenkei}, \citenamefont {Couture},\ and\ \citenamefont
  {Tanter}}]{errico2015ultrafast}%
  \BibitemOpen
  \bibfield  {author} {\bibinfo {author} {\bibfnamefont {C.}~\bibnamefont
  {Errico}}, \bibinfo {author} {\bibfnamefont {J.}~\bibnamefont {Pierre}},
  \bibinfo {author} {\bibfnamefont {S.}~\bibnamefont {Pezet}}, \bibinfo
  {author} {\bibfnamefont {Y.}~\bibnamefont {Desailly}}, \bibinfo {author}
  {\bibfnamefont {Z.}~\bibnamefont {Lenkei}}, \bibinfo {author} {\bibfnamefont
  {O.}~\bibnamefont {Couture}}, \ and\ \bibinfo {author} {\bibfnamefont
  {M.}~\bibnamefont {Tanter}},\ }\bibfield  {title} {\enquote {\bibinfo {title}
  {Ultrafast ultrasound localization microscopy for deep super-resolution
  vascular imaging},}\ }\href {\doibase 10.1038/nature16066} {\bibfield
  {journal} {\bibinfo  {journal} {Nat.}\ }\textbf {\bibinfo {volume} {527}},\
  \bibinfo {pages} {499--502} (\bibinfo {year} {2015})}\BibitemShut {NoStop}%
\bibitem [{\citenamefont {Tachibana}\ and\ \citenamefont
  {Tachibana}(1999)}]{1347-4065-38-5S-3014}%
  \BibitemOpen
  \bibfield  {author} {\bibinfo {author} {\bibfnamefont {K.}~\bibnamefont
  {Tachibana}}\ and\ \bibinfo {author} {\bibfnamefont {S.}~\bibnamefont
  {Tachibana}},\ }\bibfield  {title} {\enquote {\bibinfo {title} {Application
  of ultrasound energy as a new drug delivery system},}\ }\href {\doibase
  10.1143/jjap.38.3014} {\bibfield  {journal} {\bibinfo  {journal} {Jpn. J.
  Appl. Phys.}\ }\textbf {\bibinfo {volume} {38}},\ \bibinfo {pages} {3014}
  (\bibinfo {year} {1999})}\BibitemShut {NoStop}%
\bibitem [{\citenamefont {Tsutsui}, \citenamefont {Xie},\ and\ \citenamefont
  {Porter}(2004)}]{Tsutsui2004}%
  \BibitemOpen
  \bibfield  {author} {\bibinfo {author} {\bibfnamefont {J.~M.}\ \bibnamefont
  {Tsutsui}}, \bibinfo {author} {\bibfnamefont {F.}~\bibnamefont {Xie}}, \ and\
  \bibinfo {author} {\bibfnamefont {R.~T.}\ \bibnamefont {Porter}},\ }\bibfield
   {title} {\enquote {\bibinfo {title} {The use of microbubbles to target drug
  delivery},}\ }\href {\doibase 10.1186/1476-7120-2-23} {\bibfield  {journal}
  {\bibinfo  {journal} {Cardiovasc. Ultrasound}\ }\textbf {\bibinfo {volume}
  {2}} (\bibinfo {year} {2004}),\ 10.1186/1476-7120-2-23}\BibitemShut {NoStop}%
\bibitem [{\citenamefont {Hernot}\ and\ \citenamefont
  {Klibanov}(2008)}]{HERNOT20081153}%
  \BibitemOpen
  \bibfield  {author} {\bibinfo {author} {\bibfnamefont {S.}~\bibnamefont
  {Hernot}}\ and\ \bibinfo {author} {\bibfnamefont {A.~L.}\ \bibnamefont
  {Klibanov}},\ }\bibfield  {title} {\enquote {\bibinfo {title} {Microbubbles
  in ultrasound-triggered drug and gene delivery},}\ }\href {\doibase
  https://doi.org/10.1016/j.addr.2008.03.005} {\bibfield  {journal} {\bibinfo
  {journal} {Adv. Drug Delivery Rev.}\ }\textbf {\bibinfo {volume} {60}},\
  \bibinfo {pages} {1153 -- 1166} (\bibinfo {year} {2008})}\BibitemShut
  {NoStop}%
\bibitem [{\citenamefont {Hynynen}(2008)}]{HYNYNEN20081209}%
  \BibitemOpen
  \bibfield  {author} {\bibinfo {author} {\bibfnamefont {K.}~\bibnamefont
  {Hynynen}},\ }\bibfield  {title} {\enquote {\bibinfo {title} {Ultrasound for
  drug and gene delivery to the brain},}\ }\href {\doibase
  https://doi.org/10.1016/j.addr.2008.03.010} {\bibfield  {journal} {\bibinfo
  {journal} {Adv. Drug Delivery Rev.}\ }\textbf {\bibinfo {volume} {60}},\
  \bibinfo {pages} {1209 -- 1217} (\bibinfo {year} {2008})}\BibitemShut
  {NoStop}%
\bibitem [{\citenamefont {Kooiman}\ \emph {et~al.}(2014)\citenamefont
  {Kooiman}, \citenamefont {Vos}, \citenamefont {Versluis},\ and\ \citenamefont
  {de~Jong}}]{KOOIMAN201428}%
  \BibitemOpen
  \bibfield  {author} {\bibinfo {author} {\bibfnamefont {K.}~\bibnamefont
  {Kooiman}}, \bibinfo {author} {\bibfnamefont {H.~J.}\ \bibnamefont {Vos}},
  \bibinfo {author} {\bibfnamefont {M.}~\bibnamefont {Versluis}}, \ and\
  \bibinfo {author} {\bibfnamefont {N.}~\bibnamefont {de~Jong}},\ }\bibfield
  {title} {\enquote {\bibinfo {title} {Acoustic behavior of microbubbles and
  implications for drug delivery},}\ }\href {\doibase
  https://doi.org/10.1016/j.addr.2014.03.003} {\bibfield  {journal} {\bibinfo
  {journal} {Adv. Drug Delivery Rev.}\ }\textbf {\bibinfo {volume} {72}},\
  \bibinfo {pages} {28 -- 48} (\bibinfo {year} {2014})}\BibitemShut {NoStop}%
\bibitem [{\citenamefont {Suslick}(1990)}]{Suslick1990}%
  \BibitemOpen
  \bibfield  {author} {\bibinfo {author} {\bibfnamefont {K.~S.}\ \bibnamefont
  {Suslick}},\ }\bibfield  {title} {\enquote {\bibinfo {title}
  {Sonochemistry},}\ }\href {\doibase 10.1126/science.247.4949.1439} {\bibfield
   {journal} {\bibinfo  {journal} {Sci.}\ }\textbf {\bibinfo {volume} {247}},\
  \bibinfo {pages} {1439--1445} (\bibinfo {year} {1990})}\BibitemShut {NoStop}%
\bibitem [{\citenamefont {Blake}(1999)}]{Blake1999}%
  \BibitemOpen
  \bibfield  {author} {\bibinfo {author} {\bibfnamefont {J.~R.}\ \bibnamefont
  {Blake}},\ }\bibfield  {title} {\enquote {\bibinfo {title} {Preface to
  acoustic cavitation and sonoluminescence},}\ }\href {\doibase
  10.1098/rsta.1999.0323} {\bibfield  {journal} {\bibinfo  {journal} {Philos.
  Trans. R. Soc. London, Ser. A}\ }\textbf {\bibinfo {volume} {357}},\ \bibinfo
  {pages} {201--201} (\bibinfo {year} {1999})}\BibitemShut {NoStop}%
\bibitem [{\citenamefont {Agarwal}\ \emph {et~al.}(2012)\citenamefont
  {Agarwal}, \citenamefont {Xu}, \citenamefont {Ng},\ and\ \citenamefont
  {Liu}}]{Agarwal2012}%
  \BibitemOpen
  \bibfield  {author} {\bibinfo {author} {\bibfnamefont {A.}~\bibnamefont
  {Agarwal}}, \bibinfo {author} {\bibfnamefont {H.}~\bibnamefont {Xu}},
  \bibinfo {author} {\bibfnamefont {W.~J.}\ \bibnamefont {Ng}}, \ and\ \bibinfo
  {author} {\bibfnamefont {Y.}~\bibnamefont {Liu}},\ }\bibfield  {title}
  {\enquote {\bibinfo {title} {Biofilm detachment by self-collapsing air
  microbubbles: a potential chemical-free cleaning technology for membrane
  biofouling},}\ }\href {\doibase 10.1039/c1jm14439a} {\bibfield  {journal}
  {\bibinfo  {journal} {J. Mater. Chem.}\ }\textbf {\bibinfo {volume} {22}},\
  \bibinfo {pages} {2203--2207} (\bibinfo {year} {2012})}\BibitemShut {NoStop}%
\bibitem [{\citenamefont {Seo}\ \emph {et~al.}(2018)\citenamefont {Seo},
  \citenamefont {Leong}, \citenamefont {Park}, \citenamefont {Hong},
  \citenamefont {Chu}, \citenamefont {Park}, \citenamefont {Kim}, \citenamefont
  {Deng}, \citenamefont {Dushnov}, \citenamefont {Soh}, \citenamefont {Rogers},
  \citenamefont {Yang},\ and\ \citenamefont {Kong}}]{Seo2018}%
  \BibitemOpen
  \bibfield  {author} {\bibinfo {author} {\bibfnamefont {Y.}~\bibnamefont
  {Seo}}, \bibinfo {author} {\bibfnamefont {J.}~\bibnamefont {Leong}}, \bibinfo
  {author} {\bibfnamefont {J.~D.}\ \bibnamefont {Park}}, \bibinfo {author}
  {\bibfnamefont {Y.~T.}\ \bibnamefont {Hong}}, \bibinfo {author}
  {\bibfnamefont {S.~H.}\ \bibnamefont {Chu}}, \bibinfo {author} {\bibfnamefont
  {C.}~\bibnamefont {Park}}, \bibinfo {author} {\bibfnamefont {D.~H.}\
  \bibnamefont {Kim}}, \bibinfo {author} {\bibfnamefont {Y.~H.}\ \bibnamefont
  {Deng}}, \bibinfo {author} {\bibfnamefont {V.}~\bibnamefont {Dushnov}},
  \bibinfo {author} {\bibfnamefont {J.}~\bibnamefont {Soh}}, \bibinfo {author}
  {\bibfnamefont {S.}~\bibnamefont {Rogers}}, \bibinfo {author} {\bibfnamefont
  {Y.~Y.}\ \bibnamefont {Yang}}, \ and\ \bibinfo {author} {\bibfnamefont
  {H.}~\bibnamefont {Kong}},\ }\bibfield  {title} {\enquote {\bibinfo {title}
  {Diatom microbubbler for active biofilm removal in confined spaces},}\ }\href
  {\doibase 10.1021/acsami.8b08643} {\bibfield  {journal} {\bibinfo  {journal}
  {ACS Appl. Mater. Interfaces}\ }\textbf {\bibinfo {volume} {10}},\ \bibinfo
  {pages} {35685--35692} (\bibinfo {year} {2018})}\BibitemShut {NoStop}%
\bibitem [{\citenamefont {Ohl}\ \emph {et~al.}(2006)\citenamefont {Ohl},
  \citenamefont {Arora}, \citenamefont {Dijkink}, \citenamefont {Janve},\ and\
  \citenamefont {Lohse}}]{Ohl2006}%
  \BibitemOpen
  \bibfield  {author} {\bibinfo {author} {\bibfnamefont {C.~D.}\ \bibnamefont
  {Ohl}}, \bibinfo {author} {\bibfnamefont {M.}~\bibnamefont {Arora}}, \bibinfo
  {author} {\bibfnamefont {R.}~\bibnamefont {Dijkink}}, \bibinfo {author}
  {\bibfnamefont {V.}~\bibnamefont {Janve}}, \ and\ \bibinfo {author}
  {\bibfnamefont {D.}~\bibnamefont {Lohse}},\ }\bibfield  {title} {\enquote
  {\bibinfo {title} {Surface cleaning from laser-induced cavitation bubbles},}\
  }\href {\doibase 10.1063/1.2337506} {\bibfield  {journal} {\bibinfo
  {journal} {Appl. Phys. Lett.}\ }\textbf {\bibinfo {volume} {89}},\ \bibinfo
  {pages} {074102} (\bibinfo {year} {2006})}\BibitemShut {NoStop}%
\bibitem [{\citenamefont {Reuter}\ \emph {et~al.}(2017)\citenamefont {Reuter},
  \citenamefont {Lauterborn}, \citenamefont {Mettin},\ and\ \citenamefont
  {Lauterborn}}]{REUTER2017542}%
  \BibitemOpen
  \bibfield  {author} {\bibinfo {author} {\bibfnamefont {F.}~\bibnamefont
  {Reuter}}, \bibinfo {author} {\bibfnamefont {S.}~\bibnamefont {Lauterborn}},
  \bibinfo {author} {\bibfnamefont {R.}~\bibnamefont {Mettin}}, \ and\ \bibinfo
  {author} {\bibfnamefont {W.}~\bibnamefont {Lauterborn}},\ }\bibfield  {title}
  {\enquote {\bibinfo {title} {Membrane cleaning with ultrasonically driven
  bubbles},}\ }\href {\doibase https://doi.org/10.1016/j.ultsonch.2016.12.012}
  {\bibfield  {journal} {\bibinfo  {journal} {Ultrason. Sonochem.}\ }\textbf
  {\bibinfo {volume} {37}},\ \bibinfo {pages} {542 -- 560} (\bibinfo {year}
  {2017})}\BibitemShut {NoStop}%
\bibitem [{\citenamefont {Lee}\ \emph {et~al.}(2015)\citenamefont {Lee},
  \citenamefont {Lee}, \citenamefont {Lee},\ and\ \citenamefont
  {Park}}]{lee2015stabilization}%
  \BibitemOpen
  \bibfield  {author} {\bibinfo {author} {\bibfnamefont {M.}~\bibnamefont
  {Lee}}, \bibinfo {author} {\bibfnamefont {E.~Y.}\ \bibnamefont {Lee}},
  \bibinfo {author} {\bibfnamefont {D.}~\bibnamefont {Lee}}, \ and\ \bibinfo
  {author} {\bibfnamefont {B.~J.}\ \bibnamefont {Park}},\ }\bibfield  {title}
  {\enquote {\bibinfo {title} {Stabilization and fabrication of microbubbles:
  applications for medical purposes and functional materials},}\ }\href@noop {}
  {\bibfield  {journal} {\bibinfo  {journal} {Soft matter}\ }\textbf {\bibinfo
  {volume} {11}},\ \bibinfo {pages} {2067--2079} (\bibinfo {year}
  {2015})}\BibitemShut {NoStop}%
\bibitem [{\citenamefont {de~Jong}\ \emph {et~al.}(1992)\citenamefont
  {de~Jong}, \citenamefont {Hoff}, \citenamefont {Skotland},\ and\
  \citenamefont {Bom}}]{DEJONG199295}%
  \BibitemOpen
  \bibfield  {author} {\bibinfo {author} {\bibfnamefont {N.}~\bibnamefont
  {de~Jong}}, \bibinfo {author} {\bibfnamefont {L.}~\bibnamefont {Hoff}},
  \bibinfo {author} {\bibfnamefont {T.}~\bibnamefont {Skotland}}, \ and\
  \bibinfo {author} {\bibfnamefont {N.}~\bibnamefont {Bom}},\ }\bibfield
  {title} {\enquote {\bibinfo {title} {Absorption and scatter of encapsulated
  gas filled microspheres: Theoretical considerations and some measurements},}\
  }\href {\doibase https://doi.org/10.1016/0041-624X(92)90041-J} {\bibfield
  {journal} {\bibinfo  {journal} {Ultrasonics}\ }\textbf {\bibinfo {volume}
  {30}},\ \bibinfo {pages} {95 -- 103} (\bibinfo {year} {1992})}\BibitemShut
  {NoStop}%
\bibitem [{\citenamefont {de~Jong}\ and\ \citenamefont
  {Hoff}(1993)}]{DEJONG1993175}%
  \BibitemOpen
  \bibfield  {author} {\bibinfo {author} {\bibfnamefont {N.}~\bibnamefont
  {de~Jong}}\ and\ \bibinfo {author} {\bibfnamefont {L.}~\bibnamefont {Hoff}},\
  }\bibfield  {title} {\enquote {\bibinfo {title} {Ultrasound scattering
  properties of albunex microspheres},}\ }\href {\doibase
  https://doi.org/10.1016/0041-624X(93)90004-J} {\bibfield  {journal} {\bibinfo
   {journal} {Ultrasonics}\ }\textbf {\bibinfo {volume} {31}},\ \bibinfo
  {pages} {175 -- 181} (\bibinfo {year} {1993})}\BibitemShut {NoStop}%
\bibitem [{\citenamefont {Church}(1995)}]{Church1995}%
  \BibitemOpen
  \bibfield  {author} {\bibinfo {author} {\bibfnamefont {C.~C.}\ \bibnamefont
  {Church}},\ }\bibfield  {title} {\enquote {\bibinfo {title} {The effects of
  an elastic solid surface layer on the radial pulsations of gas bubbles},}\
  }\href {\doibase 10.1121/1.412091} {\bibfield  {journal} {\bibinfo  {journal}
  {J. Acoust. Soc. Am.}\ }\textbf {\bibinfo {volume} {97}},\ \bibinfo {pages}
  {1510--1521} (\bibinfo {year} {1995})}\BibitemShut {NoStop}%
\bibitem [{\citenamefont {Brenner}, \citenamefont {Lohse},\ and\ \citenamefont
  {Dupont}(1995)}]{brenner1995bubble}%
  \BibitemOpen
  \bibfield  {author} {\bibinfo {author} {\bibfnamefont {M.~P.}\ \bibnamefont
  {Brenner}}, \bibinfo {author} {\bibfnamefont {D.}~\bibnamefont {Lohse}}, \
  and\ \bibinfo {author} {\bibfnamefont {T.~F.}\ \bibnamefont {Dupont}},\
  }\bibfield  {title} {\enquote {\bibinfo {title} {Bubble shape oscillations
  and the onset of sonoluminescence},}\ }\href {\doibase
  10.1103/PhysRevLett.75.954} {\bibfield  {journal} {\bibinfo  {journal} {Phys.
  Rev. Lett.}\ }\textbf {\bibinfo {volume} {75}},\ \bibinfo {pages} {954}
  (\bibinfo {year} {1995})}\BibitemShut {NoStop}%
\bibitem [{\citenamefont {Hao}\ and\ \citenamefont
  {Prosperetti}(1999)}]{doi:10.1063/1.869996}%
  \BibitemOpen
  \bibfield  {author} {\bibinfo {author} {\bibfnamefont {Y.}~\bibnamefont
  {Hao}}\ and\ \bibinfo {author} {\bibfnamefont {A.}~\bibnamefont
  {Prosperetti}},\ }\bibfield  {title} {\enquote {\bibinfo {title} {The effect
  of viscosity on the spherical stability of oscillating gas bubbles},}\ }\href
  {\doibase 10.1063/1.869996} {\bibfield  {journal} {\bibinfo  {journal} {Phys.
  Fluids}\ }\textbf {\bibinfo {volume} {11}},\ \bibinfo {pages} {1309--1317}
  (\bibinfo {year} {1999})}\BibitemShut {NoStop}%
\bibitem [{\citenamefont {{Versluis}}\ \emph {et~al.}(2004)\citenamefont
  {{Versluis}}, \citenamefont {{van der Meer}}, \citenamefont {{Lohse}},
  \citenamefont {{Palanchon}}, \citenamefont {{Goertz}}, \citenamefont
  {{Chin}},\ and\ \citenamefont {{de Jong}}}]{1417703}%
  \BibitemOpen
  \bibfield  {author} {\bibinfo {author} {\bibfnamefont {M.}~\bibnamefont
  {{Versluis}}}, \bibinfo {author} {\bibfnamefont {S.~M.}\ \bibnamefont {{van
  der Meer}}}, \bibinfo {author} {\bibfnamefont {D.}~\bibnamefont {{Lohse}}},
  \bibinfo {author} {\bibfnamefont {P.}~\bibnamefont {{Palanchon}}}, \bibinfo
  {author} {\bibfnamefont {D.}~\bibnamefont {{Goertz}}}, \bibinfo {author}
  {\bibfnamefont {C.~T.}\ \bibnamefont {{Chin}}}, \ and\ \bibinfo {author}
  {\bibfnamefont {N.}~\bibnamefont {{de Jong}}},\ }\bibfield  {title} {\enquote
  {\bibinfo {title} {Microbubble surface modes [ultrasound contrast agents]},}\
  }in\ \href {\doibase 10.1109/ULTSYM.2004.1417703} {\emph {\bibinfo
  {booktitle} {IEEE Ultrasonics Symposium, 2004}}},\ Vol.~\bibinfo {volume}
  {1}\ (\bibinfo {year} {2004})\ pp.\ \bibinfo {pages} {207--209
  Vol.1}\BibitemShut {NoStop}%
\bibitem [{\citenamefont {{van der Meer}}\ \emph {et~al.}(2006)\citenamefont
  {{van der Meer}}, \citenamefont {{Dollet}}, \citenamefont {{Goertz}},
  \citenamefont {{de Jong}}, \citenamefont {{Versluis}},\ and\ \citenamefont
  {{Lohse}}}]{4151897}%
  \BibitemOpen
  \bibfield  {author} {\bibinfo {author} {\bibfnamefont {S.~M.}\ \bibnamefont
  {{van der Meer}}}, \bibinfo {author} {\bibfnamefont {B.}~\bibnamefont
  {{Dollet}}}, \bibinfo {author} {\bibfnamefont {D.~E.}\ \bibnamefont
  {{Goertz}}}, \bibinfo {author} {\bibfnamefont {N.}~\bibnamefont {{de Jong}}},
  \bibinfo {author} {\bibfnamefont {M.}~\bibnamefont {{Versluis}}}, \ and\
  \bibinfo {author} {\bibfnamefont {D.}~\bibnamefont {{Lohse}}},\ }\bibfield
  {title} {\enquote {\bibinfo {title} {Surface modes of ultrasound contrast
  agent microbubbles},}\ }in\ \href {\doibase 10.1109/ULTSYM.2006.41} {\emph
  {\bibinfo {booktitle} {2006 IEEE Ultrasonics Symposium}}}\ (\bibinfo {year}
  {2006})\ pp.\ \bibinfo {pages} {112--115}\BibitemShut {NoStop}%
\bibitem [{\citenamefont {Dollet}\ \emph {et~al.}(2008)\citenamefont {Dollet},
  \citenamefont {van~der Meer}, \citenamefont {Garbin}, \citenamefont
  {de~Jong}, \citenamefont {Lohse},\ and\ \citenamefont
  {Versluis}}]{dollet2008nonspherical}%
  \BibitemOpen
  \bibfield  {author} {\bibinfo {author} {\bibfnamefont {B.}~\bibnamefont
  {Dollet}}, \bibinfo {author} {\bibfnamefont {S.~M.}\ \bibnamefont {van~der
  Meer}}, \bibinfo {author} {\bibfnamefont {V.}~\bibnamefont {Garbin}},
  \bibinfo {author} {\bibfnamefont {N.}~\bibnamefont {de~Jong}}, \bibinfo
  {author} {\bibfnamefont {D.}~\bibnamefont {Lohse}}, \ and\ \bibinfo {author}
  {\bibfnamefont {M.}~\bibnamefont {Versluis}},\ }\bibfield  {title} {\enquote
  {\bibinfo {title} {Nonspherical oscillations of ultrasound contrast agent
  microbubbles},}\ }\href {\doibase
  https://doi.org/10.1016/j.ultrasmedbio.2008.01.020} {\bibfield  {journal}
  {\bibinfo  {journal} {Ultrasound Med. Biol.}\ }\textbf {\bibinfo {volume}
  {34}},\ \bibinfo {pages} {1465--1473} (\bibinfo {year} {2008})}\BibitemShut
  {NoStop}%
\bibitem [{\citenamefont {Vos}\ \emph {et~al.}(2011)\citenamefont {Vos},
  \citenamefont {Dollet}, \citenamefont {Versluis},\ and\ \citenamefont
  {de~Jong}}]{vos2011nonspherical}%
  \BibitemOpen
  \bibfield  {author} {\bibinfo {author} {\bibfnamefont {H.~J.}\ \bibnamefont
  {Vos}}, \bibinfo {author} {\bibfnamefont {B.}~\bibnamefont {Dollet}},
  \bibinfo {author} {\bibfnamefont {M.}~\bibnamefont {Versluis}}, \ and\
  \bibinfo {author} {\bibfnamefont {N.}~\bibnamefont {de~Jong}},\ }\bibfield
  {title} {\enquote {\bibinfo {title} {Nonspherical shape oscillations of
  coated microbubbles in contact with a wall},}\ }\href {\doibase
  https://doi.org/10.1016/j.ultrasmedbio.2011.02.013} {\bibfield  {journal}
  {\bibinfo  {journal} {Ultrasound Med. Biol.}\ }\textbf {\bibinfo {volume}
  {37}},\ \bibinfo {pages} {935--948} (\bibinfo {year} {2011})}\BibitemShut
  {NoStop}%
\bibitem [{\citenamefont {Liu}\ \emph {et~al.}(2011)\citenamefont {Liu},
  \citenamefont {Sugiyama}, \citenamefont {Takagi},\ and\ \citenamefont
  {Matsumoto}}]{liu2011numerical}%
  \BibitemOpen
  \bibfield  {author} {\bibinfo {author} {\bibfnamefont {Y.}~\bibnamefont
  {Liu}}, \bibinfo {author} {\bibfnamefont {K.}~\bibnamefont {Sugiyama}},
  \bibinfo {author} {\bibfnamefont {S.}~\bibnamefont {Takagi}}, \ and\ \bibinfo
  {author} {\bibfnamefont {Y.}~\bibnamefont {Matsumoto}},\ }\bibfield  {title}
  {\enquote {\bibinfo {title} {Numerical study on the shape oscillation of an
  encapsulated microbubble in ultrasound field},}\ }\href {\doibase
  10.1063/1.3578493} {\bibfield  {journal} {\bibinfo  {journal} {Phys. Fluids}\
  }\textbf {\bibinfo {volume} {23}},\ \bibinfo {pages} {041904} (\bibinfo
  {year} {2011})}\BibitemShut {NoStop}%
\bibitem [{\citenamefont {Tsiglifis}\ and\ \citenamefont
  {Pelekasis}(2011)}]{tsiglifis2011parametric}%
  \BibitemOpen
  \bibfield  {author} {\bibinfo {author} {\bibfnamefont {K.}~\bibnamefont
  {Tsiglifis}}\ and\ \bibinfo {author} {\bibfnamefont {N.~A.}\ \bibnamefont
  {Pelekasis}},\ }\bibfield  {title} {\enquote {\bibinfo {title} {Parametric
  stability and dynamic buckling of an encapsulated microbubble subject to
  acoustic disturbances},}\ }\href {\doibase 10.1063/1.3536646} {\bibfield
  {journal} {\bibinfo  {journal} {Phys. Fluids}\ }\textbf {\bibinfo {volume}
  {23}},\ \bibinfo {pages} {012102} (\bibinfo {year} {2011})}\BibitemShut
  {NoStop}%
\bibitem [{\citenamefont {Tamadapu}, \citenamefont {Grishenkov},\ and\
  \citenamefont {Eriksson}(2016)}]{tamadapu2016modeling}%
  \BibitemOpen
  \bibfield  {author} {\bibinfo {author} {\bibfnamefont {G.}~\bibnamefont
  {Tamadapu}}, \bibinfo {author} {\bibfnamefont {D.}~\bibnamefont
  {Grishenkov}}, \ and\ \bibinfo {author} {\bibfnamefont {A.}~\bibnamefont
  {Eriksson}},\ }\bibfield  {title} {\enquote {\bibinfo {title} {Modeling and
  parametric investigation of thick encapsulated microbubble's nonspherical
  oscillations},}\ }\href {\doibase 10.1121/1.4967737} {\bibfield  {journal}
  {\bibinfo  {journal} {J. Acoust. Soc. Am.}\ }\textbf {\bibinfo {volume}
  {140}},\ \bibinfo {pages} {3884--3895} (\bibinfo {year} {2016})}\BibitemShut
  {NoStop}%
\bibitem [{\citenamefont {Hilgenfeldt}, \citenamefont {Lohse},\ and\
  \citenamefont {Brenner}(1996)}]{doi:10.1063/1.869131}%
  \BibitemOpen
  \bibfield  {author} {\bibinfo {author} {\bibfnamefont {S.}~\bibnamefont
  {Hilgenfeldt}}, \bibinfo {author} {\bibfnamefont {D.}~\bibnamefont {Lohse}},
  \ and\ \bibinfo {author} {\bibfnamefont {M.~P.}\ \bibnamefont {Brenner}},\
  }\bibfield  {title} {\enquote {\bibinfo {title} {Phase diagrams for
  sonoluminescing bubbles},}\ }\href {\doibase 10.1063/1.869131} {\bibfield
  {journal} {\bibinfo  {journal} {Phys. Fluids}\ }\textbf {\bibinfo {volume}
  {8}},\ \bibinfo {pages} {2808--2826} (\bibinfo {year} {1996})}\BibitemShut
  {NoStop}%
\bibitem [{\citenamefont {Brenner}, \citenamefont {Hilgenfeldt},\ and\
  \citenamefont {Lohse}(2002)}]{RevModPhys.74.425}%
  \BibitemOpen
  \bibfield  {author} {\bibinfo {author} {\bibfnamefont {M.~P.}\ \bibnamefont
  {Brenner}}, \bibinfo {author} {\bibfnamefont {S.}~\bibnamefont
  {Hilgenfeldt}}, \ and\ \bibinfo {author} {\bibfnamefont {D.}~\bibnamefont
  {Lohse}},\ }\bibfield  {title} {\enquote {\bibinfo {title} {Single-bubble
  sonoluminescence},}\ }\href {\doibase 10.1103/RevModPhys.74.425} {\bibfield
  {journal} {\bibinfo  {journal} {Rev. Mod. Phys.}\ }\textbf {\bibinfo {volume}
  {74}},\ \bibinfo {pages} {425--484} (\bibinfo {year} {2002})}\BibitemShut
  {NoStop}%
\bibitem [{\citenamefont {Loughran}, \citenamefont {Eckersley},\ and\
  \citenamefont {Tang}(2012)}]{doi:10.1121/1.4707479}%
  \BibitemOpen
  \bibfield  {author} {\bibinfo {author} {\bibfnamefont {J.}~\bibnamefont
  {Loughran}}, \bibinfo {author} {\bibfnamefont {R.~J.}\ \bibnamefont
  {Eckersley}}, \ and\ \bibinfo {author} {\bibfnamefont {M.-X.}\ \bibnamefont
  {Tang}},\ }\bibfield  {title} {\enquote {\bibinfo {title} {Modeling
  non-spherical oscillations and stability of acoustically driven shelled
  microbubbles},}\ }\href {\doibase 10.1121/1.4707479} {\bibfield  {journal}
  {\bibinfo  {journal} {J. Acoust. Soc. Am.}\ }\textbf {\bibinfo {volume}
  {131}},\ \bibinfo {pages} {4349--4357} (\bibinfo {year} {2012})}\BibitemShut
  {NoStop}%
\bibitem [{\citenamefont {Doinikov}(2004)}]{doinikov2004translational}%
  \BibitemOpen
  \bibfield  {author} {\bibinfo {author} {\bibfnamefont {A.~A.}\ \bibnamefont
  {Doinikov}},\ }\bibfield  {title} {\enquote {\bibinfo {title} {Translational
  motion of a bubble undergoing shape oscillations},}\ }\href {\doibase
  doi:10.1017/S0022112003006220} {\bibfield  {journal} {\bibinfo  {journal} {J.
  Fluid Mech.}\ }\textbf {\bibinfo {volume} {501}},\ \bibinfo {pages} {1--24}
  (\bibinfo {year} {2004})}\BibitemShut {NoStop}%
\bibitem [{\citenamefont {Shaw}(2006)}]{Shaw2006}%
  \BibitemOpen
  \bibfield  {author} {\bibinfo {author} {\bibfnamefont {S.~J.}\ \bibnamefont
  {Shaw}},\ }\bibfield  {title} {\enquote {\bibinfo {title} {Translation and
  oscillation of a bubble under axisymmetric deformation},}\ }\href {\doibase
  10.1063/1.2227047} {\bibfield  {journal} {\bibinfo  {journal} {Phys. Fluids}\
  }\textbf {\bibinfo {volume} {18}},\ \bibinfo {pages} {072104} (\bibinfo
  {year} {2006})}\BibitemShut {NoStop}%
\bibitem [{\citenamefont {Shaw}(2009)}]{shaw2009stability}%
  \BibitemOpen
  \bibfield  {author} {\bibinfo {author} {\bibfnamefont {S.}~\bibnamefont
  {Shaw}},\ }\bibfield  {title} {\enquote {\bibinfo {title} {The stability of a
  bubble in a weakly viscous liquid subject to an acoustic traveling wave},}\
  }\href {\doibase 10.1063/1.3076932} {\bibfield  {journal} {\bibinfo
  {journal} {Phys. Fluids}\ }\textbf {\bibinfo {volume} {21}},\ \bibinfo
  {pages} {022104} (\bibinfo {year} {2009})}\BibitemShut {NoStop}%
\bibitem [{\citenamefont {Shaw}(2017)}]{shaw2017nonspherical}%
  \BibitemOpen
  \bibfield  {author} {\bibinfo {author} {\bibfnamefont {S.~J.}\ \bibnamefont
  {Shaw}},\ }\bibfield  {title} {\enquote {\bibinfo {title} {Nonspherical
  sub-millimeter gas bubble oscillations: Parametric forcing and nonlinear
  shape mode coupling},}\ }\href {\doibase 10.1063/1.5005599} {\bibfield
  {journal} {\bibinfo  {journal} {Phys. Fluids}\ }\textbf {\bibinfo {volume}
  {29}},\ \bibinfo {pages} {122103} (\bibinfo {year} {2017})}\BibitemShut
  {NoStop}%
\bibitem [{\citenamefont {Gu{\'e}dra}\ and\ \citenamefont
  {Inserra}(2018)}]{guedra2018bubble}%
  \BibitemOpen
  \bibfield  {author} {\bibinfo {author} {\bibfnamefont {M.}~\bibnamefont
  {Gu{\'e}dra}}\ and\ \bibinfo {author} {\bibfnamefont {C.}~\bibnamefont
  {Inserra}},\ }\bibfield  {title} {\enquote {\bibinfo {title} {Bubble shape
  oscillations of finite amplitude},}\ }\href {\doibase 10.1017/jfm.2018.768}
  {\bibfield  {journal} {\bibinfo  {journal} {J. Fluid Mech.}\ }\textbf
  {\bibinfo {volume} {857}},\ \bibinfo {pages} {681--703} (\bibinfo {year}
  {2018})}\BibitemShut {NoStop}%
\bibitem [{\citenamefont {Dash}\ and\ \citenamefont
  {Tamadapu}(2022{\natexlab{a}})}]{dash_tamadapu_2022}%
  \BibitemOpen
  \bibfield  {author} {\bibinfo {author} {\bibfnamefont {N.}~\bibnamefont
  {Dash}}\ and\ \bibinfo {author} {\bibfnamefont {G.}~\bibnamefont
  {Tamadapu}},\ }\bibfield  {title} {\enquote {\bibinfo {title} {Radial
  dynamics of an encapsulated microbubble with interface energy},}\ }\href
  {\doibase 10.1017/jfm.2021.979} {\bibfield  {journal} {\bibinfo  {journal}
  {J. Fluid Mech.}\ }\textbf {\bibinfo {volume} {932}},\ \bibinfo {pages} {A26}
  (\bibinfo {year} {2022}{\natexlab{a}})}\BibitemShut {NoStop}%
\bibitem [{\citenamefont {Steigmann}\ and\ \citenamefont
  {Ogden}(1999)}]{steigmann}%
  \BibitemOpen
  \bibfield  {author} {\bibinfo {author} {\bibfnamefont {D.~J.}\ \bibnamefont
  {Steigmann}}\ and\ \bibinfo {author} {\bibfnamefont {R.~W.}\ \bibnamefont
  {Ogden}},\ }\bibfield  {title} {\enquote {\bibinfo {title} {Elastic
  surface-substrate interactions},}\ }\href {\doibase 10.1098/rspa.1999.0320}
  {\bibfield  {journal} {\bibinfo  {journal} {Proc. R. Soc. London, Ser. A}\
  }\textbf {\bibinfo {volume} {455}},\ \bibinfo {pages} {437--474} (\bibinfo
  {year} {1999})}\BibitemShut {NoStop}%
\bibitem [{\citenamefont {Gao}\ \emph {et~al.}(2014)\citenamefont {Gao},
  \citenamefont {Huang}, \citenamefont {Qu},\ and\ \citenamefont
  {Fang}}]{GAO201459}%
  \BibitemOpen
  \bibfield  {author} {\bibinfo {author} {\bibfnamefont {X.}~\bibnamefont
  {Gao}}, \bibinfo {author} {\bibfnamefont {Z.}~\bibnamefont {Huang}}, \bibinfo
  {author} {\bibfnamefont {J.}~\bibnamefont {Qu}}, \ and\ \bibinfo {author}
  {\bibfnamefont {D.}~\bibnamefont {Fang}},\ }\bibfield  {title} {\enquote
  {\bibinfo {title} {A curvature-dependent interfacial energy-based interface
  stress theory and its applications to nano-structured materials: (i) general
  theory},}\ }\href {\doibase https://doi.org/10.1016/j.jmps.2014.01.010}
  {\bibfield  {journal} {\bibinfo  {journal} {J. Mech. Phys. Solids}\ }\textbf
  {\bibinfo {volume} {66}},\ \bibinfo {pages} {59 -- 77} (\bibinfo {year}
  {2014})}\BibitemShut {NoStop}%
\bibitem [{\citenamefont {Cattaneo}\ and\ \citenamefont
  {Supponen}(2023)}]{D3SM00871A}%
  \BibitemOpen
  \bibfield  {author} {\bibinfo {author} {\bibfnamefont {M.}~\bibnamefont
  {Cattaneo}}\ and\ \bibinfo {author} {\bibfnamefont {O.}~\bibnamefont
  {Supponen}},\ }\bibfield  {title} {\enquote {\bibinfo {title} {Shell
  viscosity estimation of lipid-coated microbubbles},}\ }\href {\doibase
  10.1039/D3SM00871A} {\bibfield  {journal} {\bibinfo  {journal} {Soft Matter}\
  }\textbf {\bibinfo {volume} {19}},\ \bibinfo {pages} {5925--5941} (\bibinfo
  {year} {2023})}\BibitemShut {NoStop}%
\bibitem [{\citenamefont {Dash}\ and\ \citenamefont
  {Tamadapu}(2022{\natexlab{b}})}]{dash2022describing}%
  \BibitemOpen
  \bibfield  {author} {\bibinfo {author} {\bibfnamefont {N.}~\bibnamefont
  {Dash}}\ and\ \bibinfo {author} {\bibfnamefont {G.}~\bibnamefont
  {Tamadapu}},\ }\bibfield  {title} {\enquote {\bibinfo {title} {Describing the
  dynamics of a nonlinear viscoelastic shelled microbubble with an interface
  energy model},}\ }\href {\doibase 10.1063/5.0127399} {\bibfield  {journal}
  {\bibinfo  {journal} {J. Appl. Phys.}\ }\textbf {\bibinfo {volume} {132}},\
  \bibinfo {pages} {204702} (\bibinfo {year} {2022}{\natexlab{b}})}\BibitemShut
  {NoStop}%
\bibitem [{\citenamefont {Fung}(1977)}]{yuan-chengfung1977}%
  \BibitemOpen
  \bibfield  {author} {\bibinfo {author} {\bibfnamefont {Y.~C.}\ \bibnamefont
  {Fung}},\ }\href@noop {} {\emph {\bibinfo {title} {A First Course in
  Continuum Mechanics}}}\ (\bibinfo  {publisher} {Prentice Hall},\ \bibinfo
  {year} {1977})\BibitemShut {NoStop}%
\bibitem [{\citenamefont {Nayfeh}(2008)}]{nayfeh2008perturbation}%
  \BibitemOpen
  \bibfield  {author} {\bibinfo {author} {\bibfnamefont {A.~H.}\ \bibnamefont
  {Nayfeh}},\ }\href@noop {} {\emph {\bibinfo {title} {Perturbation methods}}}\
  (\bibinfo  {publisher} {John Wiley \& Sons},\ \bibinfo {year}
  {2008})\BibitemShut {NoStop}%
\bibitem [{\citenamefont {Allen}\ and\ \citenamefont
  {Roy}(2000{\natexlab{a}})}]{allen2000dynamics}%
  \BibitemOpen
  \bibfield  {author} {\bibinfo {author} {\bibfnamefont {J.~S.}\ \bibnamefont
  {Allen}}\ and\ \bibinfo {author} {\bibfnamefont {R.~A.}\ \bibnamefont
  {Roy}},\ }\bibfield  {title} {\enquote {\bibinfo {title} {Dynamics of gas
  bubbles in viscoelastic fluids. ii. nonlinear viscoelasticity},}\ }\href
  {\doibase https://doi.org/10.1121/1.1289361} {\bibfield  {journal} {\bibinfo
  {journal} {J. Acoust. Soc. Am.}\ }\textbf {\bibinfo {volume} {108}},\
  \bibinfo {pages} {1640--1650} (\bibinfo {year}
  {2000}{\natexlab{a}})}\BibitemShut {NoStop}%
\bibitem [{\citenamefont {Allen}\ and\ \citenamefont
  {Roy}(2000{\natexlab{b}})}]{doi:10.1121/1.429344}%
  \BibitemOpen
  \bibfield  {author} {\bibinfo {author} {\bibfnamefont {J.~S.}\ \bibnamefont
  {Allen}}\ and\ \bibinfo {author} {\bibfnamefont {R.~A.}\ \bibnamefont
  {Roy}},\ }\bibfield  {title} {\enquote {\bibinfo {title} {Dynamics of gas
  bubbles in viscoelastic fluids. i. linear viscoelasticity},}\ }\href
  {\doibase 10.1121/1.429344} {\bibfield  {journal} {\bibinfo  {journal} {J.
  Acoust. Soc. Am.}\ }\textbf {\bibinfo {volume} {107}},\ \bibinfo {pages}
  {3167--3178} (\bibinfo {year} {2000}{\natexlab{b}})}\BibitemShut {NoStop}%
\bibitem [{\citenamefont {Doinikov}, \citenamefont {Zhao},\ and\ \citenamefont
  {Dayton}(2009)}]{doinikov2009modeling}%
  \BibitemOpen
  \bibfield  {author} {\bibinfo {author} {\bibfnamefont {A.~A.}\ \bibnamefont
  {Doinikov}}, \bibinfo {author} {\bibfnamefont {S.}~\bibnamefont {Zhao}}, \
  and\ \bibinfo {author} {\bibfnamefont {P.~A.}\ \bibnamefont {Dayton}},\
  }\bibfield  {title} {\enquote {\bibinfo {title} {Modeling of the acoustic
  response from contrast agent microbubbles near a rigid wall},}\ }\href
  {\doibase 10.1016/j.ultras.2008.07.017} {\bibfield  {journal} {\bibinfo
  {journal} {Ultrasonics}\ }\textbf {\bibinfo {volume} {49}},\ \bibinfo {pages}
  {195--201} (\bibinfo {year} {2009})}\BibitemShut {NoStop}%
\bibitem [{\citenamefont {Suslov}, \citenamefont {Ooi},\ and\ \citenamefont
  {Manasseh}(2012)}]{suslov2012nonlinear}%
  \BibitemOpen
  \bibfield  {author} {\bibinfo {author} {\bibfnamefont {S.~A.}\ \bibnamefont
  {Suslov}}, \bibinfo {author} {\bibfnamefont {A.}~\bibnamefont {Ooi}}, \ and\
  \bibinfo {author} {\bibfnamefont {R.}~\bibnamefont {Manasseh}},\ }\bibfield
  {title} {\enquote {\bibinfo {title} {Nonlinear dynamic behavior of
  microscopic bubbles near a rigid wall},}\ }\href {\doibase
  10.1103/PhysRevE.85.066309} {\bibfield  {journal} {\bibinfo  {journal} {Phys.
  Rev. E}\ }\textbf {\bibinfo {volume} {85}},\ \bibinfo {pages} {066309}
  (\bibinfo {year} {2012})}\BibitemShut {NoStop}%
\bibitem [{\citenamefont {Stride}\ \emph {et~al.}(2009)\citenamefont {Stride},
  \citenamefont {Porter}, \citenamefont {Prieto},\ and\ \citenamefont
  {Pankhurst}}]{stride2009enhancement}%
  \BibitemOpen
  \bibfield  {author} {\bibinfo {author} {\bibfnamefont {E.}~\bibnamefont
  {Stride}}, \bibinfo {author} {\bibfnamefont {C.}~\bibnamefont {Porter}},
  \bibinfo {author} {\bibfnamefont {A.~G.}\ \bibnamefont {Prieto}}, \ and\
  \bibinfo {author} {\bibfnamefont {Q.}~\bibnamefont {Pankhurst}},\ }\bibfield
  {title} {\enquote {\bibinfo {title} {Enhancement of microbubble mediated gene
  delivery by simultaneous exposure to ultrasonic and magnetic fields},}\
  }\href {\doibase https://doi.org/10.1016/j.ultrasmedbio.2008.11.010}
  {\bibfield  {journal} {\bibinfo  {journal} {Ultrasound Med. Biol.}\ }\textbf
  {\bibinfo {volume} {35}},\ \bibinfo {pages} {861--868} (\bibinfo {year}
  {2009})}\BibitemShut {NoStop}%
\bibitem [{\citenamefont {Jamburidze}\ \emph {et~al.}(2019)\citenamefont
  {Jamburidze}, \citenamefont {Huerre}, \citenamefont {Baresch}, \citenamefont
  {Poulichet}, \citenamefont {De~Corato},\ and\ \citenamefont
  {Garbin}}]{jamburidze2019nanoparticle}%
  \BibitemOpen
  \bibfield  {author} {\bibinfo {author} {\bibfnamefont {A.}~\bibnamefont
  {Jamburidze}}, \bibinfo {author} {\bibfnamefont {A.}~\bibnamefont {Huerre}},
  \bibinfo {author} {\bibfnamefont {D.}~\bibnamefont {Baresch}}, \bibinfo
  {author} {\bibfnamefont {V.}~\bibnamefont {Poulichet}}, \bibinfo {author}
  {\bibfnamefont {M.}~\bibnamefont {De~Corato}}, \ and\ \bibinfo {author}
  {\bibfnamefont {V.}~\bibnamefont {Garbin}},\ }\bibfield  {title} {\enquote
  {\bibinfo {title} {Nanoparticle-coated microbubbles for combined ultrasound
  imaging and drug delivery},}\ }\href {\doibase
  https://doi.org/10.1021/acs.langmuir.8b04008} {\bibfield  {journal} {\bibinfo
   {journal} {Langmuir}\ }\textbf {\bibinfo {volume} {35}},\ \bibinfo {pages}
  {10087--10096} (\bibinfo {year} {2019})}\BibitemShut {NoStop}%
\bibitem [{\citenamefont {Malvar}, \citenamefont {Gontijo},\ and\ \citenamefont
  {Cunha}(2018)}]{malvar2018nonlinear}%
  \BibitemOpen
  \bibfield  {author} {\bibinfo {author} {\bibfnamefont {S.}~\bibnamefont
  {Malvar}}, \bibinfo {author} {\bibfnamefont {R.}~\bibnamefont {Gontijo}}, \
  and\ \bibinfo {author} {\bibfnamefont {F.}~\bibnamefont {Cunha}},\ }\bibfield
   {title} {\enquote {\bibinfo {title} {Nonlinear motion of an oscillating
  bubble immersed in a magnetic fluid},}\ }\href {\doibase
  https://link.springer.com/article/10.1007/s10665-017-9917-7} {\bibfield
  {journal} {\bibinfo  {journal} {J. Eng. Math.}\ }\textbf {\bibinfo {volume}
  {108}},\ \bibinfo {pages} {143--170} (\bibinfo {year} {2018})}\BibitemShut
  {NoStop}%
\end{thebibliography}%





\end{document}